\documentclass[%
 preprint,
 amsmath,amssymb,
 aps,
]{revtex4-2}

\usepackage{graphicx}
\usepackage{dcolumn}
\usepackage{bm}
\usepackage{color}
\usepackage[dvipsnames]{xcolor}
\usepackage{booktabs}

\begin{document}


\title{Comparison of regularized SCAN functional with SCAN functional with and without self-interaction
 for a wide-array of properties}

\author{Yoh Yamamoto$^*$}
\author{Alan Salcedo$^*$}
\author{Carlos M. Diaz$^{*\S}$}
\author{Md Shamsul Alam$^{*\S}$}
\author{Tunna Baruah$^{*\S}$}
\author{Rajendra R. Zope$^{*\S{a)}}$}
\affiliation{$^*$Department of Physics, The University of Texas at El Paso, El Paso, Texas, 79968}
\affiliation{$^{\S}$Computational Science Program, The University of Texas at El Paso, El Paso, Texas, 79968}
\email{$^{a)}$}

\date{\today}

\begin{abstract}
The Strongly Constrained and Appropriately Normed (SCAN) functional is a non-empirical meta-generalized-gradient approximation (meta-GGA) functional that satisfies all the known constraints that a meta-GGA functional can, but it also exhibits a great degree of sensitivity to numerical grids. Its numerical complexities are amplified when used in Perdew-Zunger (PZ) self-interaction correction (SIC) which requires evaluating energies and potentials using orbital densities that vary far more rapidly than spin densities. Recent regularization of the SCAN functional (rSCAN) simplifies numerical complexities of SCAN at the expense of violation of some exact constraints. To develop a good understanding of the performance of rSCAN and the effect of loss of an exact constraint at the limit of slowly varying density, we have compared its performance against SCAN for vibrational frequencies, infra-red and Raman intensities of water clusters, electric dipole moments, spin magnetic moments of a few molecular magnets, weak interaction energies of dimers, barrier heights of reactions, and atomization energies for benchmark sets of molecules. Likewise, we examined the performance of SIC-rSCAN using the PZ-SIC method by studying atomic total energies, ionization potentials and electron affinities, molecular atomization energies, barrier heights, and dissociation and reaction energies. We find that rSCAN requires a much less dense numerical grid and gives very similar results as SCAN for all properties examined with the exception of atomization energies which are somewhat worse in rSCAN. On the other hand, SIC-rSCAN gives marginally better performance than SIC-SCAN for almost all properties studied in this work. 
\end{abstract}

\maketitle

\section{\label{sec:introduction}Introduction}
The Kohn-Sham (KS) formulation of density functional theory (DFT)\cite{PhysRev.140.A1133} is the dominant quantum mechanical approach for materials 
simulations. The accuracy of (DFT) calculations depends on the approximation used for the exchange-correlation (XC) energy term. Meta-generalized gradient approximations (meta-GGA) to the XC functionals are placed at the third rung of the Jacob's ladder of density functionals\cite{doi:10.1063/1.1390175} and have a mathematical form given as
\begin{equation}\label{eq:Exc}
    E_{XC}[\rho_\uparrow,\rho_\downarrow]=\int d\vec{r}\rho(\vec{r}) \epsilon_{XC}(\rho_\uparrow,\rho_\downarrow,\vec{\nabla}\rho_\uparrow,\vec{\nabla}\rho_\downarrow,\tau_\uparrow,\tau_\downarrow)
\end{equation}
where $\rho(\vec{r})$ is the electron density, and $\tau$ is the kinetic energy density 
typically defined as
\begin{equation}
    \tau_\sigma = \frac 1 2 \sum_i^{occ} \vec{\nabla} \psi_{i\sigma} \cdot \vec{\nabla} \psi_{i\sigma}
\end{equation}
where $\sigma$ is the spin index and the summation $i$ runs over occupied orbitals. In general, the meta-GGA XC functionals provide better chemical accuracy in comparison to the local spin density approximation (LSDA) and generalized gradient approximations (GGA). For many properties, it provides results comparable to or better than hybrid functionals\cite{doi:10.1063/1.464304,jaramillo2003local} that include a fraction of Hartree-Fock exchange. The Tao-Perdew-Staroverov-Scuseria (TPSS)\cite{PhysRevLett.91.146401,doi:10.1063/1.1665298} (appeared in 2003) and Minnesota M06L\cite{doi:10.1063/1.2370993,zhao2008m06} (2006) functionals are two widely used examples of meta-GGAs, and these, in general, show better performance than the Perdew-Burke-Ernzerhof (PBE)\cite{PhysRevLett.77.3865,PhysRevLett.78.1396} GGA.

In 2015, Sun, Ruzsinszky, and Perdew reported a new meta-GGA functional which they called the ``Strongly Constrained and Appropriately Normed (SCAN)" functional\cite{PhysRevLett.115.036402}. The SCAN functional is designed to satisfy all 17 known constraints of a semilocal functional. SCAN performs well for total energies of atoms and molecules, atomization energies, and short range Van der Waals (vdW) interactions. SCAN is also self-correlation free. In SCAN, the kinetic energy density $\tau$ is used to construct an iso-orbital indicator $\alpha$ defined as
\begin{equation}
    \alpha = \frac{\tau - \tau^W}{\tau^{unif}} > 0
\end{equation}
where $\tau^W = |\vec{\nabla} \rho|^2/8 \rho$ is the Weizs\"acker kinetic energy density\cite{weizsacker1935theorie} and $\tau^{unif}=(3/10)(3\pi^2)^{2/3}\rho^{5/3}$ is the kinetic energy density in the uniform density limit. Spin index is omitted here and in the remainder of the text for simplicity. The iso-orbital indicator quantity is used to identify different bond types (covalent, metallic, and weak bonds). Interpolation between $\alpha=0$ and $1$ and extrapolation to $\alpha\gg 1$ within the functional provide a means to satisfy some of the exact constraints. In contrast, revTPSS\cite{PhysRevLett.103.026403,PhysRevLett.106.179902} uses $z=\tau^W/\tau$ and M06L uses $t^{-1}=\tau/\tau^{unif}$ to differentiate different orbital-overlap regions\cite{PhysRevLett.111.106401}. 

SCAN has been successfully used to study several properties of materials in the last five years though there are also reports of its failure\cite{PhysRevLett.121.207201}. SCAN shows systematic improvement in predicting electronic structure properties of thin film and layered materials\cite{buda2017characterization}. The functional is also capable of describing the molecular bond types and characteristics as accurately as hybrid functionals\cite{sun2016accurate} and even better in some cases. Chen \textit{et al.}\cite{Chen10846} applied SCAN to study liquid water and found that SCAN can accurately describe its structural, electronic, and dynamic properties. Tran \textit{et al.}\cite{doi:10.1063/1.4948636} performed extensive tests on the lattice constants, bulk moduli, and cohesive energies for rung 1-4 functionals and reported that SCAN is one of the most accurate functionals for predicting those properties among them. Yang \textit{et al.}\cite{PhysRevB.100.035132}, in order to understand the better structure prediction accuracy of SCAN functional, studied the relationship between coordination environments, the description of attractive vdW interactions, and the ground-state prediction in bulk main-group solids. They noted that unlike PBE, the SCAN functional is free from systematic under-coordination error. They further concluded that the medium-range vdW interaction is correctly described in SCAN.

The success of SCAN has also led to several derivatives of the functional. SCAN+rVV10\cite{PhysRevX.6.041005} supplements the accurate short- and intermediate-range vdW interactions of SCAN with the long-range vdW from rVV10\cite{PhysRevB.87.041108} and shows promising performance in layered materials. Hybrid and double-hybrid functionals based on SCAN were proposed by Hui and Chai (SCAN0, SCAN0-DH, SCAN-QIDH, and SCAN0-2)\cite{doi:10.1063/1.4940734}. Mezei {\it et al.} proposed revSCAN\cite{doi:10.1021/acs.jctc.8b00072} where they modified SCAN by revising the form of its correlation part and found improved single-orbital electron densities and atomization energies. This group further extended revSCAN with the nonlocal VV10 dispersion-correction (revSCANVV10) and its global hybrid with 25\% exact exchange (revSCAN0)\cite{doi:10.1021/acs.jctc.8b00072}. There is also a deorbitalized (Laplacian-dependent) version of SCAN called SCAN-L\cite{PhysRevB.98.115161} where the explicit orbital dependent quantity $\alpha[\rho]$ in SCAN is replaced with a Laplacian-dependent quantity. SCAN-L showed success for applications in extended systems with a speedup up to a factor of 3. 

Meta-GGA functionals are semi-local and, in principle, are computationally more efficient than hybrid functionals which include a certain percentage of Hartree-Fock exchange. It has however been found that the implementation of meta-GGA functionals is usually more difficult and pose numerical challenges due to the need for very dense numerical grids. This has been noted in a number of works\cite{grafenstein2007avoiding,grafenstein2007efficient,dasgupta2017standard,johnson2009oscillations,wheeler2010integration,PhysRevB.93.205205,doi:10.1063/1.5120532} related to implementation and the users of electronic structure codes are cautioned to be careful in appropriate grids when using the meta-GGA functional. We have recently implemented the SCAN functional in our code\cite{doi:10.1063/1.5120532} and also encountered its sensitivity to choice of numerical grid.

Recently, Furness, and Sun\cite{PhysRevB.99.041119} have examined the source of the numerical sensitivity of SCAN. They reported that the numerical problem arises from the iso-orbital indicator $\alpha$. As a proof of concept for designing a more numerically stable meta-GGA functional, they replaced the iso-orbital indicator $\alpha$ in the meta-GGA made simple 2 (MS2) functional\cite{doi:10.1063/1.4789414} with a numerically more stable $\beta$ in the MS2$\beta$ functional where $\beta$ is defined as
\begin{equation}
    \beta = \frac{\tau - \tau^W}{\tau + \tau^{unif}}.
\end{equation}
The use of $\beta$ in place of $\alpha$ leads to divergence-free XC potentials, and further development of a $\beta$-based SCAN functional is expected.

Very recently, Bart\'ok and Yates used the SCAN functional to generate a library of ultrasoft pseudopotentials\cite{PhysRevB.99.235103,doi:10.1063/1.5094646} for SCAN calculations on periodic systems. They noted severe numerical instabilities in generating the pseudopotentials. To alleviate these difficulties they proposed a modification of SCAN that replaces the problematic function in SCAN with a numerically stable polynomial function\cite{doi:10.1063/1.5094646}. The design of this regularized SCAN (rSCAN) functional has the same motivation as MS2$\beta$, that is, to make SCAN computationally more stable.

Many failures of density functional approximations (DFA) have been attributed to the self-interaction error (SIE) which imits the broad applicability of the DFAs to settings where atoms are at or near their equilibrium positions. Since SCAN has in general seemed to be broadly transferable and successful in describing a wide range of properties, our interest is in applying the Perdew-Zunger self-interaction correction (PZSIC)\cite{PhysRevB.23.5048} to SCAN to extend its range of accuracy to stretched-bond situations and calculations of properties such as chemical transition state barriers and to improve the predictive ability of SCAN.
A few methodologies have been developed to remove SIE from DFA calculations with mixing of Hartree-Fock with DFAs being the most widely used approach. The best known self-interaction correction (SIC) method to systematically eliminate SIE is the PZSIC wherein the SIE is removed on an orbital-by-orbital basis.
The PZSIC method is therefore an orbital dependent theory and requires evaluating the XC energy and potentials using orbital densities. This further increases the computational complexity as the orbital densities vary much more rapidly than the total electron (spin) densities. Accurate descriptions of XC contributions to the Hamiltonian matrix thus put stringent demand on numerical grids. We have recently used SCAN with PZSIC\cite{doi:10.1063/1.5120532} in the FLOSIC code\cite{FLOSICcode} which required adaptation\cite{doi:10.1063/1.5120532} of the variational numerical grid algorithm of Pederson and Jackson\cite{PhysRevB.41.7453}. The numerical grids required in these calculations are substantially larger than the ones needed for functionals on the lower rungs of Jacob's ladder of XC functionals. An attractive feature of the rSCAN functional of Bart\'ok and Yates is that it requires a much less dense numerical grid than SCAN. This feature is very attractive for PZSIC calculations. We have implemented the rSCAN functional in the FLOSIC code that performs the PZSIC calculations using the Fermi-L\"{o}wdin local orbitals. The present work compares the performance of SIC-rSCAN and SIC-SCAN for various properties. During the course of this work, Mej\'{i}a-Rodr\'{i}guez and Trickey\cite{doi:10.1063/1.5120408} reported the assessment of rSCAN for heat of formation, lattice parameter, and vibrational frequencies. They concluded that, while rSCAN does alleviate grid sensitivity, rSCAN and SCAN are not fully interchangeable. The focus of this work is (1) to extend the Mej\'{i}a-Rodr\'{i}guez and Trickey's comparison at the DFA level by including ionization potentials (IPs), electron affinities (EAs), weak interactions, dipole moments, in addition to vibrational frequencies and atomization energies and (2) to assess the performance of SIC-rSCAN against SIC-SCAN since the application to SIC is important.

\section{Computational method}
All calculations reported here are done using the FLOSIC code\cite{FLOSICcode} in which PZSIC is implemented using Fermi-L\"owdin orbitals (FLOs). FLOSIC is based on the UTEP-NRLMOL code and inherits all of its well-tested numerical features. These include a variational integration mesh\cite{PhysRevB.41.7453} that provides high accuracy integrals for total energies, matrix elements, and charge densities. We used the default NRLMOL Cartesian Gaussian orbital basis sets given by Porezag and Pederson\cite{PhysRevA.60.2840} in the FLOSIC code which are of roughly quadruple zeta quality\cite{doi:10.1002/jcc.25586}. For anion calculations, we included additional long range s, p, and d single Gaussian orbitals to the default NRLMOL basis to better describe the extended nature of the anionic states. 

The SCAN meta-GGA is implemented in the FLOSIC code using the approach discussed in a recent article\cite{doi:10.1063/1.5120532}. An integration by parts and Hamiltonian mixing approaches are used for the meta-GGA calculation in FLOSIC where the Hamiltonian matrix elements of the meta-GGA term are obtained as
\begin{equation}
    \int \psi_i(\vec{r}) \frac{\delta E_{XC}[\tau[\rho]]}{\delta\rho(\vec{r})} \psi_j(\vec{r})
    \approx \frac{1}{2} \int \frac{\delta E_{XC}[\tau]}{\delta\tau(\vec{r})} \vec{\nabla}\psi_i(\vec{r}) \cdot \vec{\nabla}\psi_j(\vec{r}) .
\end{equation}

We find that the default variational NRLMOL integration grid typically used for LSDA and GGA functionals is not capable of reliably capturing the shape of the meta-GGA XC potential. 
In our SCAN implementation\cite{doi:10.1063/1.5120532}, we started with a brute force approach of mesh generation where the NRLMOL variational mesh was used as a starting point and radial grid points with uniform increments were added until $E_{XC}$ converged. Subsequently, the radial mesh is adjusted such that the radial points are less dense in the non-problematic areas while maintaining the required grid density in the problematic areas. This allowed required integrals to be calculated with specified accuracy but still resulted in a very large density of grid points compared to the default variational mesh. The rSCAN functional of Bart\'ok and Yates replaces the problematic region $0 < \alpha < 2.5$ of the switching function $f(\alpha)$ used in the SCAN functional\cite{doi:10.1063/1.5094646} by a polynomial of degree 7. {\bf The specifics of this polynomial are provided in supplementary information. } In our rSCAN implementation in the FLOSIC code, we generate the mesh as mentioned above for the SCAN functional and further eliminated superfluous grid points. While the simplified rSCAN functional is designed to be far less demanding on numerical grids compared to the SCAN functional, it still requires a denser grid than many GGAs. Mej\'{i}a-Rodr\'{i}guez and Trickey have also commented\cite{doi:10.1063/1.5120408} that rSCAN mesh sensitivity is similar to SOGGA11. 
To accurately integrate the XC potential, we have designed a few different modifications into the mesh generation mechanism of the FLOSIC code for SIC-SCAN/rSCAN\cite{doi:10.1063/1.5120532}. To meet the goal of assessing the performance of rSCAN against SCAN and to examine the computational efficiency of rSCAN with respect to SCAN we adopt the following procedure. For assessment purpose we use very dense mesh (referred to mesh A hereafter) in computing all results for both SCAN and rSCAN functionals. To examine the computational efficiency of rSCAN we repeat the calculations with a mesh that is roughly $2-5$ times coarser (referred to mesh B) than mesh A. Both mesh A and mesh B gave same results using rSCAN functional.

\paragraph{FLOSIC}
Fermi-L\"owdin orbital SIC (FLOSIC) is a method for applying PZSIC to eliminate one electron SIE. The FLOSIC has been used to study ionization energies, electron affinities, exchange coupling, weekly bound anions, polarizabilities etc.\cite{C9CP06106A, Jackson_2019, PhysRevA.100.012505, doi:10.1021/acs.jpca.8b09940, doi:10.1063/1.5125205, doi:10.1002/jcc.25767, doi:10.1063/1.5050809, doi:10.1021/acs.jctc.8b00344, doi:10.1063/1.4996498}.
In PZSIC, the SIE is eliminated on an orbital by orbital basis using the following prescription,
\begin{equation}
    E^{PZSIC}[\rho_{\uparrow},\rho_{\downarrow}]
    =E^{DFA}[\rho_{\uparrow},\rho_{\downarrow}]
    - \sum_{\alpha,\sigma}^{occ}\{ U^{SIC}[\rho_{\alpha,\sigma}] + E_{XC}^{DFA}[\rho_{\alpha,\sigma},0] \}.
\end{equation}
Here, $\rho_{\alpha,\sigma}$ is the density of $\alpha^{th}$ orbital and $\sigma$ is the spin density. The orbital density is computed using the local orbitals instead of canonical KS orbitals. The local orbitals used are the FLOs where the Fermi orbitals are constructed using the Fermi orbital descriptor (FOD) positions for the transformation from KS orbitals as
\begin{equation}
    \phi_\alpha(\vec{r}) = \frac{\sum_i \psi_i(\textbf{a}_\alpha)\psi_i(\vec{r})}{\sqrt{\rho(\textbf{a}_\alpha)}}
\end{equation}
where \textbf{a}$_\alpha$ is the FOD. The Fermi orbitals are not necessarily orthogonal. Hence the L\"owdin orthogonalization\cite{doi:10.1063/1.1747632} is performed to obtain the orthonormal set of FLOs. As the local orbitals depend on the FODs, their positions have to be optimized. We have optimized the FODs for the SCAN functional and used the same set of FODs to perform self-consistent SIC-rSCAN calculations. 
To validate this choice, we performed full optimization of FODs for the six molecules from the AE6 set within SIC-rSCAN. We found that the total energy on average shifts by 61.3 $\mu$Ha per system, while the average FOD displacement is 0.033 Bohr per FOD. This shows that FODs that minimize the total energies for SIC-SCAN can be safely used in SIC-rSCAN calculations. 

Although rSCAN alleviates the numerical problem of the original SCAN functional in terms of integration grid, the problem somewhat persists in the FLOSIC calculations. The FLOSIC calculations require computing the exchange-potential dependent quantities like $E_{XC}$ or contributions from XCs to the Hamiltonian matrix using FLO densities. This results in another numerical complication. Meta-GGA functionals such as SCAN use the iso-orbital indicator $\alpha$. In the standard DFA calculations, one can evaluate $\alpha$ on grid easily using KS orbitals. The evaluation of $\alpha$ using FLOs can be rather problematic since its numerator $\tau - \tau^W$ is always close to 0, and at the same time its denominator $\tau^{unif}$ can also become very small in magnitude. This occasionally causes incorrect numerical evaluation of $\alpha \ll 0$ which can result in numerical instability (e.g. leading to an SCF convergence issue). For SIC-rSCAN calculations, when we encounter $\alpha \ll 0$, we set $F_x(\alpha) = F_x(0)$ and $dF_x(\alpha)/d\alpha = 0$. This is a fair assumption for FLOs since $\tau \approx \tau^W$, and small $\tau^{unif}$ means that $dF_x(\alpha)/d\alpha$ should vanish.

\section{Results }
\begin{figure}
    \centering
    \includegraphics[width=0.8\columnwidth,trim = {0 0 0 0}, clip]{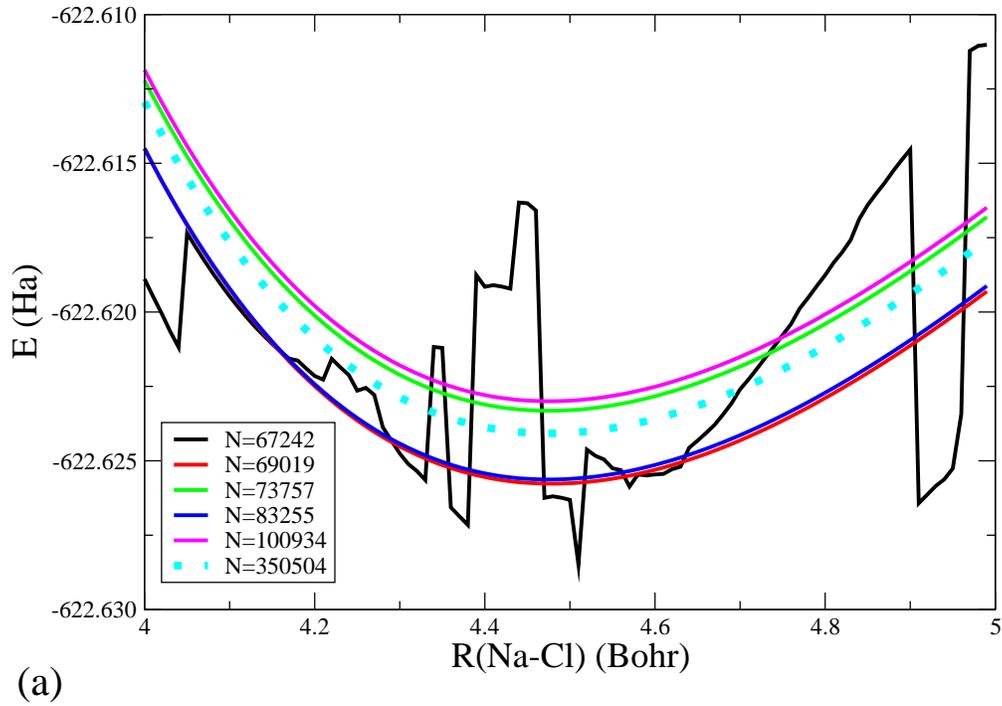}
    
    \vspace{1cm}
    
    \includegraphics[width=0.8\columnwidth,trim = {0 0 0 0}, clip]{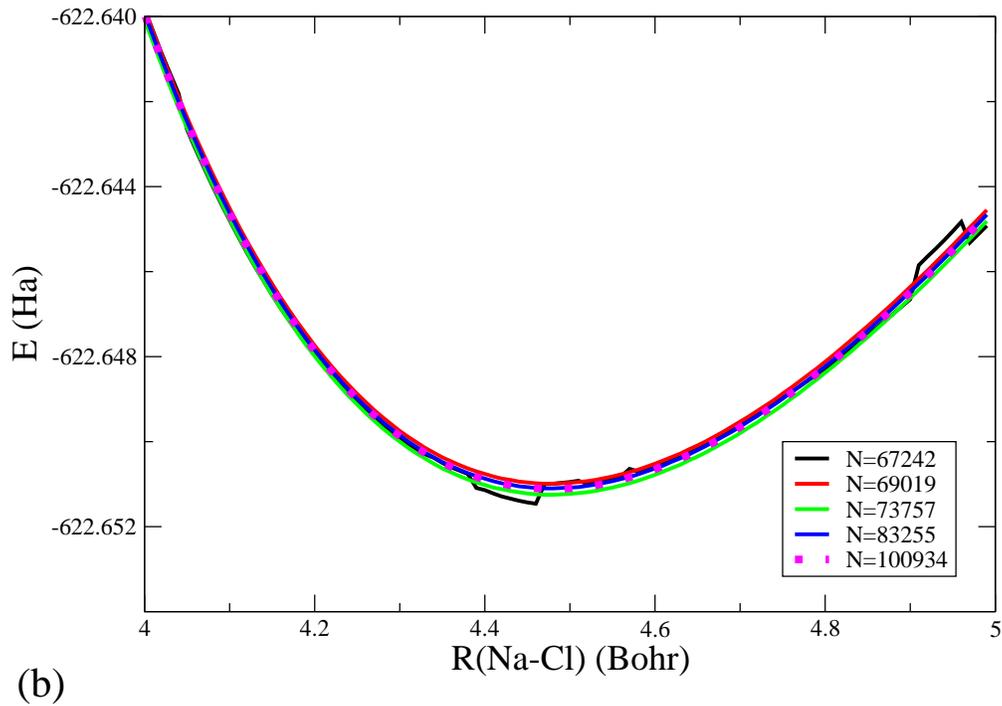}
    \caption{Energy surfaces of NaCl dimers using (a) SCAN and (b) rSCAN with various radial {\bf grid point density settings in the FLOSIC code}. {\bf $N$s shown are the averaged total grid points for a given mesh setting}. rSCAN energy curves converge faster with respect to $N$ than SCAN.}
    \label{fig:dimers}
\end{figure}

Very recently, Mej\'{i}a-Rodr\'{i}guez and Trickey\cite{doi:10.1063/1.5120408} as well as Bart\'ok and Yates\cite{doi:10.1063/1.5128484} reported the 
assessment of rSCAN for heat of formation, lattice parameter, and vibrational frequencies. Therefore, we focus here on molecular properties and on a comparison of SCAN-rSIC and SCAN-SIC. First, we consider the numerical simplification of rSCAN over SCAN functional, we computed the energy of NaCl molecule within these approximations as a function of distance for various grid densities, starting with the default variational mesh of the FLOSIC code. The grid density was then increased using the approach mentioned in previous section. The potential energy surface of the molecule as a function of bondlength is shown in Fig. \ref{fig:dimers}.

Both rSCAN and SCAN show kinks in the energy curves when the FLOSIC default mesh is used, but rSCAN showed fewer numbers of kinks. Those kinks disappear as the radial grid density is increased. As can be seen from the figure, the rSCAN total energy converges much faster with respect to the number of radial grid points $N$ than in SCAN. We note that although the kinks disappeared at grid points $N=69019$ for both the functionals, for SCAN more grid points are required to see the energy convergence than rSCAN. 

Frequencies of normal modes of vibrations are particularly sensitive to the numerical grids. We have computed the vibrational frequencies of water clusters with SCAN and rSCAN and compared them against CCSD(T) calculations by Miliordos \textit{et al.}\cite{doi:10.1063/1.4820448}. The details of calculations and vibrational frequencies are given in supplementary information. Briefly, for cluster sizes n=$1-3$, SCAN and rSCAN show comparable normal mode frequencies but show some differences for n$=4$ and $5$. The average grid points needed for vibrational frequency calculation for the (H$_2$O)$_5$ using SCAN and rSCAN are 576667 and 298307, respectively. The deviation of SCAN and rSCAN frequencies with respect to CCSD(T) increases from 10\% for water monomer to 60\% for pentamer. The rSCAN vibrational frequencies are within 10\% of SCAN frequencies (Table \ref{tab:vibrational}). In addition, we have studied infra-red (IR) spectra and Raman spectra using the two functionals (cf. Table S5-S9). The deviations of IR intensities against MP2 are shown in Table \ref{tab:IRintensities}. We find that the intensities of SCAN and rSCAN show close agreement within differences of 3\% for the active IR bands and 4\% for the majority of the active Raman bands with some exceptions. From the computational efficiency point of view, the rSCAN functional thus have significant edge over SCAN for calculations of vibrational frequencies and IR or Raman spectra.

\begin{table*}
\caption{\label{tab:vibrational}Mean absolute deviations (MADs) of harmonic vibrational frequencies of water clusters (cm$^{-1}$).}
\footnotetext{CCSD(T) values are from reference \cite{doi:10.1063/1.4820448}} 
\begin{ruledtabular}
\begin{tabular}{cccc}
Cluster & SCAN against CCSD(T)$^{\text{a}}$ & rSCAN against CCSD(T)$^{\text{a}}$ & rSCAN against SCAN\\\hline
H$_2$O  & 7.0 & 8.3 & 2.4 \\
(H$_2$O)$_2$ & 17.8 &	18.7 &	7.2\\
(H$_2$O)$_3$ & 45.9 &	50.0 &	4.8\\
(H$_2$O)$_4$ & 58.3 &	61.9 &	4.5 \\
(H$_2$O)$_5$ & 57.5 &	60.1 &	3.5 \\
\end{tabular}
\end{ruledtabular}
\end{table*}

\begin{table*}
\caption{\label{tab:IRintensities}Mean absolute deviations (MADs) of IR intensities of water clusters (Km/mol).}
\footnotetext{MP2/aug–cc–pVDZ values are from reference \cite{doi:10.1063/1.4820448}} 
\begin{ruledtabular}
\begin{tabular}{cccc}
Cluster & SCAN against MP2$^{\text{a}}$ & rSCAN against MP2$^{\text{a}}$ & rSCAN against SCAN\\\hline
H$_2$O       & 5.6  & 5.4  & 0.9 \\
(H$_2$O)$_2$ & 33.4 & 38.4 & 7.7 \\
(H$_2$O)$_3$ & 42.9 & 47.4 & 12.1\\
(H$_2$O)$_4$ & 56.1 & 54.7 & 11.8\\
(H$_2$O)$_5$ & 53.9 & 50.7 & 5.0 \\
\end{tabular}
\end{ruledtabular}
\end{table*}

\subsection{Single molecular magnet}
Recently, Fu and Singh\cite{PhysRevLett.121.207201} reported that the SCAN functional cannot describe the stability and properties of phases of Fe that are important for steel. They find that SCAN tends to overestimate magnetic energies for several elemental solids. In their subsequent work\cite{PhysRevB.100.045126}, 
they compared the performance of SCAN with other functionals for ferromagnetic Fe, Co, and Ni. They noted that the exchange splitting for open shells was enhanced when the SCAN functional was used. We use finite molecules such as single molecular magnets (SMMs) as test systems to examine performance of SCAN and rSCAN for magnetic properties. The SMMs are metal-organic compounds that exhibit magnetic hysteresis at molecular scale. They are of interest in condensed matter physics because of their potential application in magnetic memory devices and due to their possible usage for quantum information process\cite{doi:10.1002/9783527823987.vol3_c11}. 
Four SMMs studied here are:
Mn$_{12}$O$_{12}$(O$_2$CH)$_{16}$(H$_2$O)$_4$ (Mn$_{12}$ for the rest), 
Fe$_4$(OCH$_2$)$_6$(C$_4$H$_9$ON)$_6$ (Fe$_4$), 
$[$Ni(hmp)(MeOH)Cl$]_4$ (Ni$_4$), and 
Co$_4$(CH$_2$C$_5$H$_4$N)$_4$(CH$_3$OH)$_4$Cl$_4$ (Co$_4$).
Mn$_{12}$O$_{12}$ (Mn$_{12}$), the first SMM \cite{doi:10.1021/ja00015a057, doi:10.1021/ja00058a027,sessoli1993magnetic}, appeared in the early 1990's. Since then, SMMs have been studied by both theorists and experimentalists alike. 

Here, we used SCAN and rSCAN to find the most stable spin state $S$ of the above mentioned SMMs. In order to find the optimal spin state (or magnetic moment), we begin calculation with a high initial spin moment and allow the system to relax to the most stable spin state. For all four SMMs we studied here, both SCAN and rSCAN found the correct spin for all the systems, $S=10,5,4$, and $6$ for Mn$_{12}$, Fe$_4$, Ni$_4$, and Co$_4$ respectively, in agreement with experimentally reported results\cite{doi:10.1021/ja00058a027,B515980F,doi:10.1063/1.1540050,doi:10.1002/anie.200351753}.

\subsection{S22}
The S22 set\cite{B600027D} consists of weakly interacting dimers composed with C, N, O, and H atoms and is used for benchmarking non-covalent interaction energies. SCAN is able to describe weak vdW interactions and is reported to have a similar performance as M06L, a functional fitted to weak interactions in its design. The comparison against the CCSD(T)/CBS reference values from Ref. [\onlinecite{B600027D}] was made. The errors are summarized in Table \ref{tab:S22}. Mean absolute error (MAE) of SCAN is 0.90 kcal/mol and rSCAN is 0.95 kcal/mol. rSCAN agrees with SCAN within 0.05 kcal/mol. Our SCAN result differs from the MAE of Sun \textit{et al.}\cite{PhysRevLett.115.036402} by 1.5 \% possibly due to choice of different basis.

\begin{table*}
\caption{\label{tab:S22} MAE and RMSE of the S22 set of weak interactions against CCSD(T) (kcal/mol).}
\begin{ruledtabular}
\begin{tabular}{cccc}
Errors  & SCAN  & rSCAN & SCAN (Sun \textit{et al.}) \\ \hline
MAE     & 0.90	& 0.95 &  0.92 \\
RMSE    & 1.23	& 1.31 &  1.22 \\
\end{tabular}
\end{ruledtabular}
\end{table*}

\subsection{BH76}
The BH76 set \cite{doi:10.1021/jp045141s} consists of two subsets of HTBH38/08 (Hydrogen transfer) and NHTBH38/08 (non-Hydrogen transfer) and is a more comprehensive benchmark set of barrier heights than the BH6 set. NHTBH38 contains nucleophilic substitution reactions, heavy atom transfer reactions, and unimolecular and association reactions. The SCAN and rSCAN calculations were performed using the reference BH76 geometries. We compared our results against the reference values from Ref. [\onlinecite{doi:10.1021/jp045141s}] that were obtained at the W1 level of theory. The W1 method is an extrapolation of CCSD(T) in the complete basis limit, and the reference values used here include spin-orbit coupling. The results are summarized in Table \ref{tab:BH76}. The MAEs of SCAN and rSCAN are 6.74 and 6.65 kcal/mol respectively. 

\begin{table*}
\caption{\label{tab:BH76} MAE and RMSE of the BH76 set of reaction barriers against the W1 theory (kcal/mol).}
\begin{ruledtabular}
\begin{tabular}{cccc}
Errors  & SCAN  & rSCAN & SCAN (Sun \textit{et al.})\footnote{Reference [\onlinecite{PhysRevLett.115.036402}]} \\ \hline
HTBH3, MAE	& 7.59 & 7.67 & 7.76 \\
HTBH3, RMSE	& 8.02 & 8.12 & 8.20\\
\hline
NHTBH3, MAE	 & 5.88 & 5.63 & 7.74\\
NHTBH3, RMSE & 7.32	& 7.21 & 8.87\\
\hline
BH76, MAE	& 6.74 &	6.65 & 7.75 \\
BH76, RMSE	& 7.68 &	7.67 & 8.54\\

\end{tabular}
\end{ruledtabular}
\end{table*}

\subsection{Dipole moment}
Hait and Head-Gordon\cite{doi:10.1021/acs.jctc.7b01252}  recently examined the performance of 88 DFAs for prediction of dipole moments ($\bm{\mu}$) using a benchmark set of 152 molecules. We used this data set of 152 molecules to compute $\bm{\mu}$ using the SCAN and rSCAN functionals. We used same geometries as in Ref. [\onlinecite{doi:10.1021/acs.jctc.7b01252}] where most of these geometries are from experiments. MAEs and root mean square errors (RMSEs) are shown in Table \ref{tab:dipole}. Hait and Head-Gordon used the aug-pc-4\cite{doi:10.1063/1.1413524,doi:10.1063/1.1465405,doi:10.1063/1.1515484,doi:10.1021/jp068677h,doi:10.1063/1.1756866} basis set whereas our calculations used the NRLMOL basis set. It is evident from the Table that the rSCAN dipole moment are in good agreement with those predicted by the SCAN functional. The differences with respect to results of Ref. [\onlinecite{doi:10.1021/acs.jctc.7b01252}] are because of the basis set choice. Our code cannot use $f$ type basis functions used in the Ref. [\onlinecite{doi:10.1021/acs.jctc.7b01252}]
\begin{table*}
\caption{\label{tab:dipole}Dipole moments with respect to CCSD(T) (in Debye).}
\begin{ruledtabular}
\begin{tabular}{ccccc}
Errors  & SCAN  & rSCAN & SCAN (Head-Gordon)\footnote{Reference [\onlinecite{doi:10.1021/acs.jctc.7b01252}]} & Expt.\footnote{Deviation between CCSD(T) from Reference [\onlinecite{doi:10.1021/acs.jctc.7b01252}] and 80 experimental dipole moments from Reference [\onlinecite{NIST_CCCBD}].}\\ \hline
MAE     & 0.103	& 0.109 &  0.092 & 0.075\\
RMSE    & 0.173	& 0.179 &  0.147 & 0.148\\
\end{tabular}
\end{ruledtabular}
\end{table*}

\section{Results -- performance of SIC-\MakeLowercase{r}SCAN}
In this section, we discuss the performance of rSCAN with SIC for atomic total energies, IPs and EAs of atoms, atomization energies of molecules, reaction barrier heights, dissociation and reaction energies, and water cluster binding energies. The SIC-rSCAN results are compared with uncorrected rSCAN, SCAN and SIC-SCAN.

\begin{figure}[b]
\includegraphics[width=0.8\columnwidth]{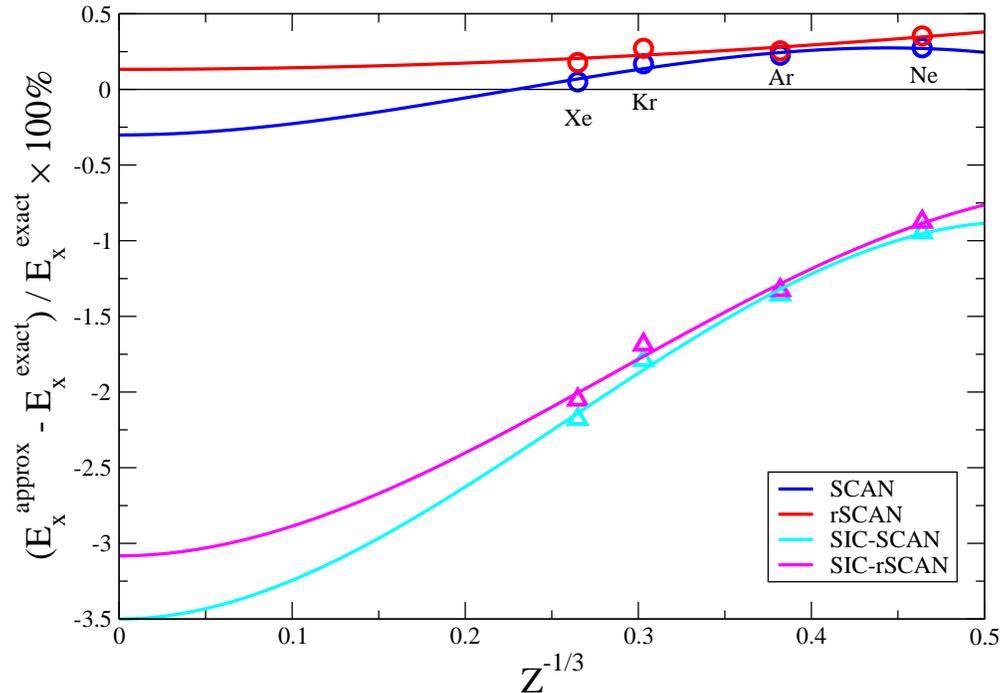}
\caption{\label{fig:largeZ} The extrapolation curves of $E_X$ to the large Z-limit for SCAN and rSCAN.}
\end{figure}

Before we discuss the energies computed using SIC-SCAN and SIC-rSCAN, we examine the behavior of $E_X$ of SIC-rSCAN in the large atomic number $Z$ limit and compare it with SCAN, rSCAN, and SIC-SCAN. Following Santra and Perdew \cite{doi:10.1063/1.5090534}, we use $f(Z)=a+b Z^{-2/3}+c Z^{-1}$ as a fitting function. 
The results are shown in Fig. \ref{fig:largeZ}. SCAN shows a small percentage error of $-0.30 \%$ in large-$Z$ limit which is a numerical artifact as SCAN is exact in the uniform gas limit. Interestingly, rSCAN also shows the large-Z limit close to $E_X^{exact}$ (percentage error, $0.13 \%$) despite the lack of exact constraint for the slowly varying density limit. Santra and Perdew have suggested the failure of SIC-SCAN in obeying the slowly varying density limit being one reason why SIC-SCAN does not perform as well as SCAN for equilibrium properties. As seen in the figure, the SIC-rSCAN curve follows SIC-SCAN but does slightly better than SIC-SCAN in the large-Z limit. 

\subsection{Atoms-  total energies}
\begin{figure}
    \centering
    \includegraphics[width=0.8\columnwidth]{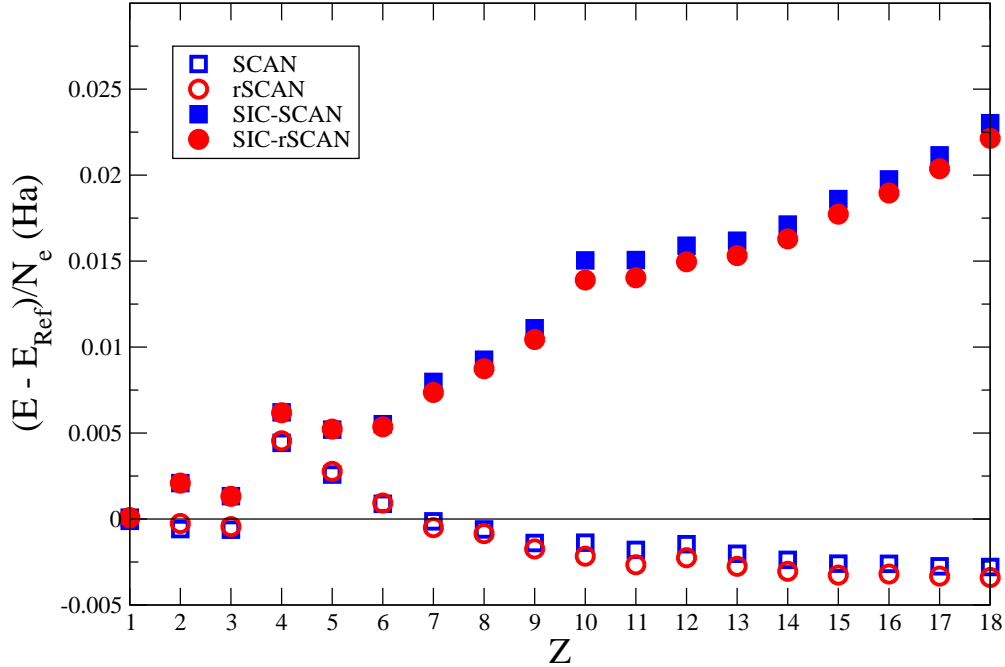}
    \caption{Total energies per electron for atoms Z$=1-18$.}
    \label{fig:atoms}
\end{figure}

We computed the total energies of atoms $Z=1-18$ and compared them against accurate non-relativistic calculations from Ref. \cite{PhysRevA.47.3649}. The results are shown in Fig. \ref{fig:atoms}. The rSCAN atomic energies have similar trend as SCAN and are slightly underestimated with respect to SCAN and experimental number. The removal of SIE using PZSIC is known to deteriorate SCAN atomic energies\cite{doi:10.1063/1.5120532,doi:10.1063/1.5087065} as the SIC is overestimated. The rSCAN results are similar to SCAN. In Table \ref{tab:atoms} we summarize the MAEs. As shown in the Table, MAEs for DFA are 0.019 and 0.027 Ha and MAEs for SIC are 0.147 and 0.140 Ha for SCAN and rSCAN respectively. The two functionals show very comparable performance. The rSCAN total energies tend to be order of $1-10$ mHa lower compared to SCAN, and correcting for SIC does not alter this trend.

\begin{table*}
\caption{\label{tab:atoms}MAE of atomic total energies in Ha.}
\footnotetext{From reference \cite{doi:10.1063/1.5120532}} 
\begin{ruledtabular}
\begin{tabular}{cc}
 Method & MAE(Ha) \\ \hline
SCAN$^\text{a}$         &   0.019 \\
rSCAN                   &   0.027\\
SIC-SCAN$^\text{a}$  &   0.147 \\
SIC-rSCAN            &   0.140 \\
\end{tabular}
\end{ruledtabular}
\end{table*}

\subsection{Atoms-  ionization potentials and electron affinities}
We computed the IPs of atoms $Z=2-36$ using $\Delta$SCF approach according to the following expression 
\begin{equation}
    E_{IP} = E_{cat} - E_{neut}.
\end{equation}
The errors in IP with respect to the experimental values from Ref. [\onlinecite{NIST_ASD}] are summarized in Table \ref{tab:ips}. At the DFA level, rSCAN shows agreement with SCAN within 0.02 eV. The SIC-SCAN and SIC-rSCAN MAEs are 0.274 and 0.222 eV for $Z=2-18$ and 0.259 and 0.342 eV for $Z=2-36$.

\begin{table*}
\caption{\label{tab:ips}MAE of $\Delta$SCF ionization potentials (eV).}
\begin{ruledtabular}
\begin{tabular}{ccc}
 Method         & Z=2-18    & Z=2-36 \\ \hline
SCAN            &   0.175   &   0.273 \\
rSCAN           &   0.193  &   0.256 \\
SIC-SCAN\footnote{From reference \cite{doi:10.1063/1.5129533}}     &  0.274    & 0.259  \\
SIC-rSCAN    &   0.222   & 0.342 \\
\end{tabular}
\end{ruledtabular}
\end{table*}

For electron affinities we considered H, Li, B, C, O, F, Na, Al, Si, P, S, Cl, K, Ti, Cu, Ga, Ge, As, Se, and Br atoms as their experimental EAs are reported in Ref. \cite{NIST_CCCBD}. Like IPs, the EAs are calculated using $\Delta$SCF. We note that for both SCAN and rSCAN the eigenvalue of the extra electron for the anions is positive indicating that it will not bind in the complete basis limit. This is a known problem in DFA\cite{PhysRevB.23.5048}. Nevertheless, we included the results for comparison as our goal is to compare SCAN against rSCAN. The MAE for SCAN with larger set is 0.148 eV compared to 0.173 eV of rSCAN. The rSCAN EAs differ from the SCAN by about $0.02-0.03$ eV depending on the data set. Application of SIC corrects the asymptotic behaviour of the potential and leads to electron binding. In this case the eigenvalue of the extra electron is negative. 
\textcolor{black}{The SIC-rSCAN errors in EAs are smaller by about $0.03-0.04$ eV indicating small improvement over the SCAN functional.}

\begin{table*}
\caption{\label{tab:eas}MAE of $\Delta$SCF electron affinities (eV).}
\begin{ruledtabular}
\begin{tabular}{ccc}
 Method         & 12 EAs    & 20 EAs \\ \hline
SCAN            &  0.115    &   0.148   \\
rSCAN           &  0.135   &   0.173  \\
SIC-SCAN\footnote{From reference \cite{doi:10.1063/1.5120532}}     &  0.364    &   0.341   \\
SIC-rSCAN    &  0.329    &   0.314 \\
\end{tabular}
\end{ruledtabular}
\end{table*}

\subsection{Atomization energies}
We used the AE6 test set to assess the performance in atomization energies. AE6 is a set of six molecules that are good representatives of the performance in atomization energies \cite{doi:10.1021/jp035287b}. It consists of SiH$_4$, SiO, S$_2$, propyne (C$_3$H$_4$), glyoxal (C$_2$H$_2$O$_2$), and cyclobutane (C$_4$H$_8$). The results are summarized in Table \ref{tab:atomization}. As has been noted in earlier works, SCAN shows a remarkable performance with MAE of only $2.85$ kcal/mol. The regularization of SCAN in rSCAN deteriorates this performance resulting in the MAE of $6.28$ kcal/mol. This failure of rSCAN was also noted in a recent study by Mej\'{i}a-Rodr\'{i}guez and Trickey\cite{doi:10.1063/1.5120408}. Bart\'ok and Yates have reassessed the standard enthalpies of formation using SCAN, rSCAN and the recent deorbitalized SCAN-L functionals\cite{doi:10.1063/1.5128484}. They concluded that the errors in enthalpies of formation are significantly larger in rSCAN compared to SCAN and SCAN-L functionals which they attributed to energy shifts of the free atom reference values. Our results are consistent with these observations. 
We note that though the MAE of rSCAN is larger than that of SCAN, it is still a substantial improvement over PBE. The MAE of rSCAN is roughly half of the MAE of PBE which is $13.43$ kcal/mol (cf. Ref. \cite{doi:10.1063/1.5129533}). The large difference between SCAN and rSCAN vanishes when SIEs are removed. In case of FLOSIC calculations, MAEs are $26.52$ and $21.63$ kcal/mol for SCAN and rSCAN respectively. As mentioned earlier the difference between SIC-SCAN and SIC-rSCAN total atomic energies is much smaller, with SIC-rSCAN being marginally better, when compared to the uncorrected SCAN and rSCAN total atomic energies. This fact, along with possible similar improvement in molecular total energies, may be the reason why SIC-rSCAN atomization energies are slightly better than their SIC-SCAN counterpart.

\begin{table*}
\caption{\label{tab:atomization}AE6 atomization energies.}
\begin{ruledtabular}
\begin{tabular}{ccc}
 Method         & MAE (kcal/mol)    &   MAPE (\%) \\ \hline
SCAN            &  2.85     &   1.15    \\
rSCAN           &  6.28     &   1.88    \\
SIC-SCAN     &  26.52    &   7.35 \\
SIC-rSCAN    &  21.63    &   6.05    \\
\end{tabular}
\end{ruledtabular}
\end{table*}

\subsection{Barrier heights}
Reaction barriers are essential chemical properties, but most DFAs fail to describe this property correctly since density functionals are primarily designed for equilibrium ground state calculations. Here, we investigate the performance of rSCAN using the BH6 benchmark set \cite{doi:10.1021/jp035287b}. BH6 consists of three chemical reactions --- 
i)   OH + CH$_4$ $\rightarrow$ CH$_3$ + H$_2$O,
ii)  H + OH $\rightarrow$ O + H$_2$, and
iii) H + H$_2$S $\rightarrow$ H$_2$ + HS.
The forward and reverse barrier heights were considered. We compute the left hand side, saddle points, and right hand sides of these chemical reactions and obtain the reaction barrier. The forward (reverse) reaction barrier is the difference of saddle point energy and the energy at the left (right) hand side of the reactions. Many DFA calculations fail to describe the barriers which require accurate prediction of energies when the bonds are stretched. The results for BH6 data sets are shown in Table \ref{tab:barrier}. From SCAN to SIC-SCAN, the MAE is reduced from $7.86$ to $2.96$ kcal/mol, and ME also shows decrease from $-7.86$ to $-0.81$ kcal/mol. Using rSCAN, MAEs are $9.41$ kcal/mol for DFA and $2.72$ kcal/mol for SIC. The rSCAN performs slightly worse than SCAN while SIC-rSCAN does marginally better than SIC-SCAN in barrier height calculations.

\begin{table*}
\caption{\label{tab:barrier}BH6 barrier heights (kcal/mol).}
\begin{ruledtabular}
\begin{tabular}{ccc}
 Method         & ME     &   MAE  \\ \hline
SCAN            &     -7.86 &  7.86     \\
rSCAN           &     -9.41 & 9.41      \\
SIC-SCAN\footnote{From reference \cite{doi:10.1063/1.5120532}}     &    -0.81 &    2.96 \\
SIC-rSCAN    &       -0.79    &   2.72    \\
\end{tabular}
\end{ruledtabular}
\end{table*}

\subsection{Dissociation and reaction energies}
Here, we calculated dissociation energies for the SIE4$\times$4 set\cite{C7CP04913G} and reaction energies for SIE11 set\cite{doi:10.1021/ct900489g}. These two test sets are part of the general main group thermochemistry, kinetics, and noncovalent interactions (GMTKN) benchmark database for studying the SIE-related problems. SIE4$\times$4 set consists of four positively charged dimers at four different separation distances $R$ which are $R/R_e=1.0, 1.25, 1.5$, and $1.75$, $R_e$ being the equilibrium distance. The dissociation energies $E_D$ are obtained as
\begin{equation}
    E_D = E(X) + E(X^+) - E(X_2^+)
\end{equation}
where $E(X_2^+)$ is the energy of the compound, $E(X)$ and $E(X^+)$ are the energies of fragments. SIE11 set consists of 5 cationic and 6 neutral reactions prone to SIEs. The MAEs with respect to the CCSD(T) reference values are summarized in Table \ref{tab:sie}. SCAN and rSCAN results with and without SIC are comparable with SCAN (SIC-rSCAN) being marginally better than rSCAN (SIC-SCAN).

\begin{table*}
\caption{\label{tab:sie}MAE of SIE4$\times$4 dissociation and SIE11 reaction energies (kcal/mol).}
\begin{ruledtabular}
\begin{tabular}{ccc}
 Method         & MAE SIE4$\times$4     &   MAE SIE11 \\ \hline
SCAN            &   17.9    &   10.1    \\
rSCAN           &   18.4    &   10.5    \\
SIC-SCAN     &   2.2     &   5.7     \\
SIC-rSCAN    &   2.1     &   5.2     \\
\end{tabular}
\end{ruledtabular}
\end{table*}

\subsection{Water hexamers}
Appropriate description of water clusters is a difficult test for the DFAs. One of many success stories of SCAN is its ability to accurately describe covalent and hydrogen bonds and vdW interactions between the water molecules\cite{Chen10846}. Water hexamers were used by Sun and coworkers to test the performance of SCAN\cite{PhysRevLett.115.036402} and also by Bart\'ok and Yates\cite{doi:10.1063/1.5094646} to test the performance of rSCAN in predicting isomer ordering.
Here we study the binding energies of water hexamers using the self-interaction-corrected rSCAN functional. The four isomers considered are as follows: the prism (P), cage (C), book (B), and ring (R), following the naming conventions used by Yagi \textit{et al.}\cite{doi:10.1021/jp802927d}. We computed the binding energies using SIC-rSCAN and compare them in Table \ref{tab:waterhex} with recent SIC-SCAN results by Sharkas \textit{et al.} \cite{FLOSIC_WATER_PNAS}. The rSCAN and SIC-rSCAN calculations show the signed errors of $-43.4$ and $-12.6$ meV/H$_2$O, respectively. These results compare well with SCAN and SIC-SCAN results of Sharkas and coworkers which has signed errors of $-41.6$ and $-13.2$ meV/H$_2$O for SCAN and SIC-SCAN, respectively. The two functionals agree within $1.8$ meV/H$_2$O.

As for the energy orderings of the water hexamer isomers, the CCSD(T) energy ordering (from the most stable to least) is shown as P $<$ C $<$ B $<$ R. It was previously shown that the SCAN functional is able to predict the same isomer ordering\cite{sun2016accurate}. FLOSIC calculations with SCAN have shown to preserve this isomer ordering\cite{FLOSIC_WATER_PNAS}. Using rSCAN, the same isomer ordering was observed both at DFA and at SIC level. These results and those in the previous sections show that rSCAN can be used in place of SCAN for SIC calculations.

\begin{table*}
\caption{\label{tab:waterhex}Signed errors in binding energies with respect to CCSD(T) for water clusters. The units are in meV/H$_2$O.}
\footnotetext{Reference \cite{FLOSIC_WATER_PNAS}}
\footnotetext{CCSD(T)-F12b reference binding energies from reference \cite{doi:10.1021/acs.jctc.6b01046}}
\begin{ruledtabular}
\begin{tabular}{cccccc}
 Cluster      & SCAN$^\text{a}$ & rSCAN & SIC-SCAN$^\text{a}$ & SIC-rSCAN & Ref.$^\text{b}$ \\ 
 \hline
(H$_2$O)$_2$   & -9.4   & -10.1   & -2.0 & -1.6 & -108.6 \\ 
(H$_2$O)$_3$   & -28.9  & -29.4   & -5.8 & -4.2 & -228.4 \\ 
(H$_2$O)$_4$   & -36.9  & -41.2   & -8.7 & -10.3 & -297.0 \\ 
(H$_2$O)$_5$   & -39.8  & -41.8   & -12.1 & -12.0 & -311.4 \\ 
 \hline
(H$_2$O)$_6$ P  & -42.4  & -43.7   &-11.7 & -10.4 & -332.4 \\
(H$_2$O)$_6$ C  & -42.7  & -44.4   &-12.0 & -11.3 & -330.5 \\
(H$_2$O)$_6$ B  & -41.7  & -43.8   &-11.3 & -11.1 & -327.3 \\
(H$_2$O)$_6$ R  & -39.4  & -41.6   &-17.6 & -17.6 & -320.1 \\
\end{tabular}
\end{ruledtabular}
\end{table*}

\section{Conclusion}
To summarize, we have implemented the recent regularized version of the SCAN functional and assessed its performance on several electronic properties ranging from atomization energies to barrier heights as well as on magnetic properties and vibrational properties. The performance appraisal was carried out for both the self-interaction corrected rSCAN functional and uncorrected rSCAN functional. The calculation of SIC energy and potentials using rSCAN can become numerically unstable due to evaluation of $\alpha$ using the FLO and FLO densities. A solution to simplify this complexity is introduced and the SIC calculations were performed for a wide array of properties. The results were compared with corresponding results using SCAN functional. Our results show that rSCAN total energies converge faster with numerical grids compared to the SCAN functional. The rSCAN results for most properties are, in general, comparable to the SCAN functional with deviation in the range $0.1-1.9$ kcal/mol. The exception is the case of atomization energies which are significantly worse compared to SCAN functional (deviation of $3.4$ kcal/mol in uncorrected DFA). For magnetic properties, we assessed uncorrected SCAN and rSCAN functionals by computing the net spin moment of a few SMMs. Our results show that both rSCAN and SCAN predict the same correct spin moment. The trends observed for uncorrected functionals are also seen when the SIEs are removed using the FLOSIC formalism. In this case for the majority of properties SIC-rSCAN results are marginally better than SIC-SCAN results. These results indicate that the impact of violation of slowly varying norm is minimal on rSCAN's performance with and without SIC. Considering that the rSCAN results are comparable to SCAN results (with exception of atomization energy), the rSCAN functional can be recommended for study with SIC as it is numerically less demanding due to need of relatively less dense numerical grids compared to the SCAN functional.

\section*{Supplementary material}
See supplementary material for detailed results of the dipole moments, IR and Raman spectra, total energies, IP, EA for the systems studied in this manuscript and detailed results for S22, BH76, AE6, BH6, SIE4$\times$4, and SIE11 molecular test sets.

\section*{Data Availability Statement}
The data that supports the findings of this study are available within the article and its supplementary material.

\begin{acknowledgments}
Authors acknowledge discussions with Dr. Jianwei Sun and Prof. John Perdew. Authors would also like to thank Prof. Koblar Jackson for comments on the manuscript. This work was supported by the US Department of Energy, Office of Science, Office of Basic Energy Sciences, as part of the Computational Chemical Sciences Program under Award No. DE-SC0018331. The work of R.R.Z. was supported in part by the US Department of Energy, Office of Science, Office of Basic Energy Sciences, under Award No. DE-SC0006818. Support for computational time at the Texas Advanced Computing Center through NSF Grant No. TG-DMR090071, and at NERSC is gratefully acknowledged.
\end{acknowledgments}

\clearpage
\bibliography{bibtex_references} 

\end{document}


\beginsupplement

\title{Supplementary information for: comparison of regularized SCAN functional with SCAN functional with and without self-interaction
for a wide-array of properties}

\author{Yoh Yamamoto$^*$}
\author{Alan Salcedo$^*$}
\author{Carlos M. Diaz$^{*\S}$}
\author{Md Shamsul Alam$^{*\S}$}   
\author{Tunna Baruah$^{*\S}$}
\author{Rajendra R. Zope$^{*\S}$}
\affiliation{$^*$Department of Physics, The University of Texas at El Paso, El Paso, Texas, 79968}
\affiliation{$^{\S}$Computational Science Program, The University of Texas at El Paso, El Paso, Texas, 79968}

{\let\vfil\relax{\let\clearpage\relax\maketitle}}

\section*{polynomial used in \MakeLowercase{r}SCAN}
\begin{figure}[b]
\includegraphics[width=0.8\columnwidth]{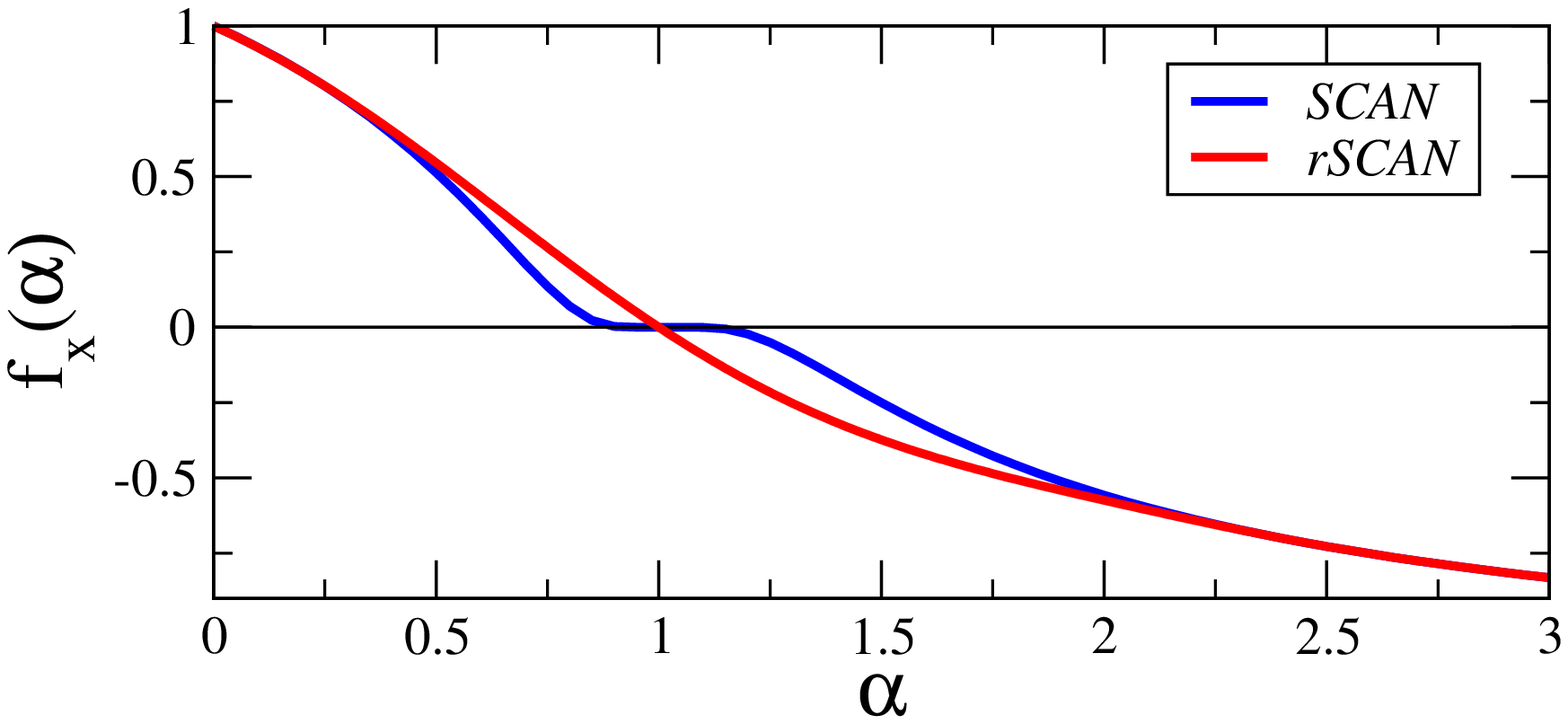}

\vspace{0.5cm}

\includegraphics[width=0.8\columnwidth,trim = {0 0 0 0}, clip]{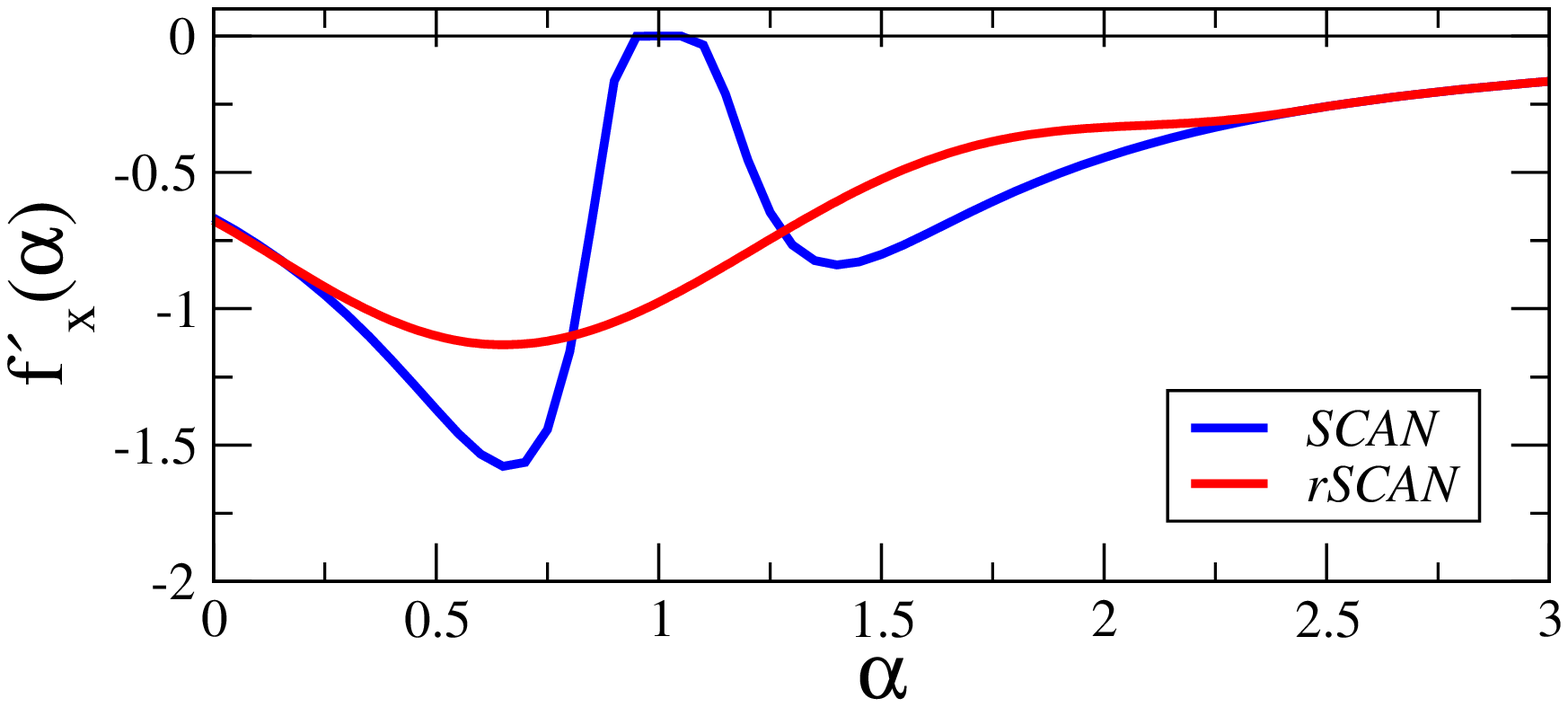}
\caption{\label{fig:switchingfunction} The switching function $f_x(\alpha)$ and $f_x'(\alpha)$ for SCAN and rSCAN.}
\end{figure}

\begin{figure}[b]
\includegraphics[width=0.8\columnwidth]{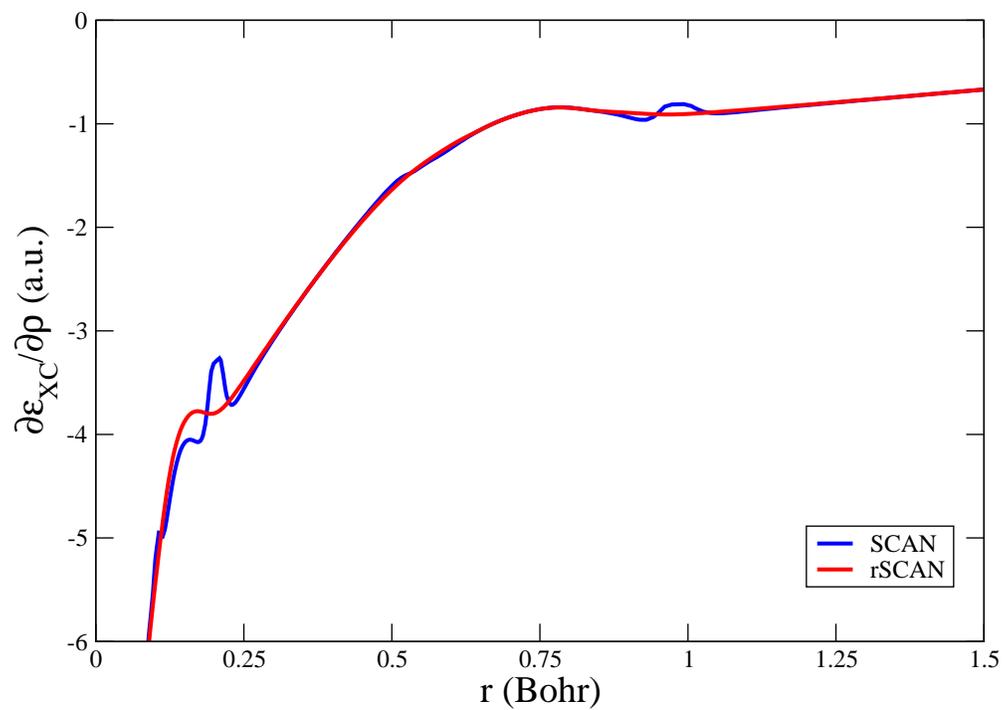}
\caption{\label{fig:scan_v_rscan}  $\partial\epsilon_{XC}/\partial\rho$ as a function of $r$ for Ar atom with SCAN and rSCAN.} 
\end{figure}

The polynomial function used in the rSCAN implementation is defined as
\begin{equation}
    f(\alpha)=c_1 + c_2 \alpha + c_3 \alpha^2 + c_4 \alpha^3 + c_5 \alpha^4 + c_6 \alpha^5 + c_7 \alpha^6 + c_8 \alpha^7
\end{equation}
for $\alpha \in [0,2.5]$ where the coefficients $c$'s are as shown in Table \ref{tab:rscancoef}. Those are the same as in  Ref. \cite{doi:10.1063/1.5094646}.
The constraints used are $f^{(0,1,2)}(0)$ and $f^{(0,1,2,3)}(2.5)$ to be identical values as the $f(\alpha)$ in SCAN at these two points. In addition, $f(1)=0$ was also used as a constraint.
The plots of $f_x(\alpha)$ and $f_x'(\alpha)$ are shown in Fig. \ref{fig:switchingfunction}.
The plot of $\partial\epsilon_{XC}/\partial\rho$ for an Ar atom is shown in Fig. \ref{fig:scan_v_rscan}.
We have also tested different choices of polynomials with the same constraints and found essentially the same results.

\begin{table*}[h]
\begin{center}
\caption{The coefficients used for the rSCAN implementation (same as Ref. \cite{doi:10.1063/1.5094646}).}
\label{tab:rscancoef}
\begin{ruledtabular}
\begin{tabular}{crr}
Coef. & Exchange   & Correlation \\ 
\colrule
$c_1$ & $ 1.000$   & $ 1.000$   \\
$c_2$ & $-0.677$   & $-0.640$   \\
$c_3$ & $-0.44456$ & $-0.4352$  \\
$c_4$ & $-0.62109$ & $-1.53568$ \\
$c_5$ & $ 1.39690$ & $ 3.06156$ \\
$c_6$ & $-0.85920$ & $-1.91571$ \\
$c_7$ & $ 0.22746$ & $ 0.51688$ \\
$c_8$ & $-0.02252$ & $-0.05185$ \\
\end{tabular}
\end{ruledtabular}
\end{center}
\end{table*}

\clearpage
\begin{table*}[h]
\begin{center}
\caption{Dipole moments $|\bm{\mu}|$ in Debye.}
\label{stab:dipolemoments}
\begin{ruledtabular}
\begin{tabular}{crrr}
System & SCAN	&	rSCAN	& CCSD(T)\footnote{Reference [\onlinecite{doi:10.1021/acs.jctc.7b01252}]} 	\\ 
\colrule
AlF	&	1.31	&	1.31	&	1.47	\\
AlH$_2$	&	0.40	&	0.40	&	0.40	\\
BeH	&	0.58	&	0.58	&	0.23	\\
BF	&	1.05	&	1.05	&	0.82	\\
BH	&	1.58	&	1.59	&	1.41	\\
BH$_2$	&	0.48	&	0.49	&	0.50	\\
BH$_2$Cl	&	0.55	&	0.55	&	0.68	\\
BH$_2$F	&	0.68	&	0.68	&	0.83	\\
BHCl$_2$	&	0.56	&	0.56	&	0.67	\\
BHF$_2$	&	0.83	&	0.83	&	0.96	\\
BN	&	2.12	&	2.06	&	2.04	\\
BO	&	2.35	&	2.33	&	2.32	\\
BS	&	0.89	&	0.85	&	0.78	\\
C$_2$H	&	0.76	&	0.74	&	0.76	\\
C$_2$H$_3$	&	0.71	&	0.70	&	0.69	\\
C$_2$H$_5$	&	0.34	&	0.34	&	0.31	\\
CF	&	0.89	&	0.90	&	0.68	\\
CF$_2$	&	0.72	&	0.72	&	0.54	\\
CH	&	1.48	&	1.48	&	1.43	\\
CH$_2$BH	&	0.56	&	0.57	&	0.62	\\
CH$_2$BOH	&	2.31	&	2.31	&	2.26	\\
CH$_2$F	&	1.27	&	1.27	&	1.38	\\
CH$_2$NH	&	2.02	&	2.00	&	2.07	\\
CH$_2$PH	&	1.02	&	0.99	&	0.87	\\
CH$_2$-singlet	&	1.83	&	1.82	&	1.49	\\
CH$_2$-triplet	&	0.59	&	0.59	&	0.59	\\
CH$_3$BH$_2$	&	0.69	&	0.70	&	0.58	\\
CH$_3$BO	&	3.77	&	3.78	&	3.68	\\
CH$_3$Cl	&	1.94	&	1.91	&	1.90	\\
CH$_3$F	&	1.72	&	1.71	&	1.81	\\
CH$_3$Li	&	5.77	&	5.76	&	5.83	\\
CH$_3$NH$_2$	&	1.36	&	1.35	&	1.39	\\
CH$_3$O	&	2.11	&	2.09	&	2.04	\\
CH$_3$OH	&	1.66	&	1.65	&	1.71	\\
CH$_3$SH	&	1.65	&	1.66	&	1.59	\\
ClCN	&	3.00	&	3.02	&	2.85	\\
ClF	&	0.82	&	0.75	&	0.88	\\
ClO$_2$	&	1.76	&	1.75	&	1.86	\\
CN	&	1.41	&	1.41	&	1.43	\\
CO	&	0.13	&	0.17	&	0.12	\\
CS	&	1.92	&	1.96	&	1.97	\\

\end{tabular}
\end{ruledtabular}
\end{center}
\end{table*} 

\begin{table*}[h]
\begin{center}
\caption*{Dipole moments continued.}
\begin{ruledtabular}
\begin{tabular}{crrr}
CSO	&	0.79	&	0.79	&	0.73	\\
FCN	&	2.33	&	2.35	&	2.18	\\
FCO	&	0.85	&	0.84	&	0.77	\\
FH-BH$_2$	&	3.03	&	3.03	&	2.97	\\
FH-NH$_2$	&	4.67	&	4.67	&	4.63	\\
FH-OH	&	3.40	&	3.40	&	3.38	\\
FNO	&	1.54	&	1.51	&	1.70	\\
H$_2$CN	&	2.52	&	2.50	&	2.49	\\
H$_2$O	&	1.86	&	1.87	&	1.86	\\
H$_2$O-Al	&	4.53	&	4.54	&	4.36	\\
H$_2$O-Cl	&	3.04	&	3.05	&	2.24	\\
H$_2$O-F	&	2.58	&	2.64	&	2.19	\\
H$_2$O-H$_2$O	&	2.78	&	2.78	&	2.73	\\
H$_2$O-Li	&	2.98	&	2.98	&	3.62	\\
H$_2$O-NH$_3$	&	3.57	&	3.58	&	3.50	\\
H$_2$S-H$_2$S	&	1.05	&	1.06	&	0.92	\\
H$_2$S-HCl	&	2.36	&	2.36	&	2.13	\\
HBH$_2$BH	&	0.85	&	0.88	&	0.84	\\
HBO	&	2.72	&	2.71	&	2.73	\\
HBS	&	1.43	&	1.40	&	1.38	\\
HCCCl	&	0.31	&	0.28	&	0.50	\\
HCCF	&	0.56	&	0.52	&	0.75	\\
HCHO	&	2.37	&	2.32	&	2.39	\\
HCHS	&	1.86	&	1.80	&	1.76	\\
HCl	&	1.16	&	1.15	&	1.11	\\
HCl-HCl	&	1.91	&	1.90	&	1.78	\\
HCN	&	3.03	&	3.02	&	3.01	\\
HCNO	&	2.60	&	2.57	&	2.96	\\
HCO	&	1.67	&	1.65	&	1.69	\\
HCOF	&	2.09	&	2.07	&	2.12	\\
HCONH$_2$	&	3.96	&	3.95	&	3.92	\\
HCOOH	&	1.48	&	1.47	&	1.38	\\
HCP	&	0.48	&	0.45	&	0.35	\\
HF	&	1.80	&	1.80	&	1.81	\\
HF-HF	&	3.42	&	3.42	&	3.40	\\
HN$_3$	&	1.77	&	1.79	&	1.66	\\
HNC	&	3.05	&	3.07	&	3.08	\\
HNCO	&	2.05	&	2.04	&	2.06	\\
HNO	&	1.57	&	1.56	&	1.65	\\
HNO$_2$	&	1.96	&	1.95	&	1.93	\\
HNS	&	1.39	&	1.39	&	1.41	\\
HO$_2$	&	2.17	&	2.21	&	2.17	\\
HOCl	&	1.55	&	1.56	&	1.52	\\
HOCN	&	3.97	&	3.99	&	3.80	\\
HOF	&	1.92	&	1.89	&	1.92	\\

\end{tabular}
\end{ruledtabular}
\end{center}
\end{table*} 

\begin{table*}[h]
\begin{center}
\caption*{Dipole moments continued.}
\begin{ruledtabular}
\begin{tabular}{crrr}
HOOH	&	1.57	&	1.57	&	1.57	\\
HPO	&	2.34	&	2.32	&	2.63	\\
LiBH$_4$	&	6.11	&	6.11	&	6.13	\\
LiCl	&	7.10	&	7.10	&	7.10	\\
LiCN	&	6.99	&	7.00	&	6.99	\\
LiF	&	6.28	&	6.28	&	6.29	\\
LiH	&	5.82	&	5.82	&	5.83	\\
LiN	&	6.84	&	6.83	&	7.06	\\
LiOH	&	4.53	&	4.53	&	4.57	\\
N$_2$H$_2$	&	2.83	&	2.83	&	2.88	\\
N$_2$H$_4$	&	2.71	&	2.71	&	2.72	\\
NaCl	&	8.85	&	8.85	&	9.01	\\
NaCN	&	8.81	&	8.82	&	8.89	\\
NaF	&	7.99	&	7.99	&	8.13	\\
NaH	&	6.33	&	6.33	&	6.40	\\
NaLi	&	0.23	&	0.25	&	0.48	\\
NaOH	&	6.63	&	6.63	&	6.77	\\
NCl	&	1.14	&	1.16	&	1.13	\\
NCO	&	0.83	&	0.84	&	0.79	\\
NF	&	0.15	&	0.18	&	0.07	\\
NF$_2$	&	0.13	&	0.11	&	0.19	\\
NH	&	1.54	&	1.54	&	1.54	\\
NH$_2$	&	1.80	&	1.80	&	1.79	\\
NH$_2$Cl	&	2.03	&	2.01	&	1.95	\\
NH$_2$F	&	2.28	&	2.24	&	2.27	\\
NH$_2$OH	&	0.70	&	0.67	&	0.70	\\
NH$_3$	&	1.55	&	1.55	&	1.53	\\
NH$_3$-BH$_3$	&	5.33	&	5.33	&	5.28	\\
NH$_3$-NH$_3$	&	2.18	&	2.19	&	2.13	\\
NH$_3$O	&	5.21	&	5.21	&	5.39	\\
NO	&	0.17	&	0.19	&	0.13	\\
NO$_2$	&	0.29	&	0.29	&	0.34	\\
NOCl	&	1.86	&	1.83	&	2.08	\\
NP	&	2.74	&	2.75	&	2.87	\\
NS	&	1.77	&	1.79	&	1.82	\\
O$_3$	&	0.63	&	0.64	&	0.57	\\
OCl	&	1.43	&	1.43	&	1.28	\\
OCl$_2$	&	0.50	&	0.48	&	0.56	\\
OF	&	0.16	&	0.19	&	0.02	\\
OF$_2$	&	0.33	&	0.31	&	0.33	\\
OH	&	1.65	&	1.65	&	1.66	\\
P$_2$H$_4$	&	1.06	&	1.08	&	1.00	\\
PCl	&	0.42	&	0.38	&	0.57	\\
PF	&	0.65	&	0.66	&	0.81	\\
PH	&	0.50	&	0.49	&	0.44	\\

\end{tabular}
\end{ruledtabular}
\end{center}
\end{table*} 

\begin{table*}[h]
\begin{center}
\caption*{Dipole moments continued.}
\begin{ruledtabular}
\begin{tabular}{crrr}
PH$_2$	&	0.62	&	0.62	&	0.55	\\
PH$_2$OH	&	1.89	&	1.92	&	0.68	\\
PH$_3$	&	0.67	&	0.69	&	0.61	\\
PH$_3$O	&	3.63	&	3.63	&	3.77	\\
PO	&	1.89	&	1.85	&	1.96	\\
PO$_2$	&	1.36	&	1.33	&	1.44	\\
PPO	&	1.74	&	1.69	&	1.88	\\
PS	&	0.55	&	0.52	&	0.68	\\
S$_2$H$_2$	&	1.19	&	1.19	&	1.14	\\
SCl	&	0.19	&	0.22	&	0.07	\\
SCl$_2$	&	0.34	&	0.31	&	0.39	\\
SF	&	0.63	&	0.59	&	0.81	\\
SF$_2$	&	0.90	&	0.87	&	1.06	\\
SH	&	0.83	&	0.83	&	0.77	\\
SH$_2$	&	1.06	&	1.07	&	0.99	\\
SiH	&	0.19	&	0.20	&	0.11	\\
SiH$_3$Cl	&	1.28	&	1.28	&	1.36	\\
SiH$_3$F	&	1.23	&	1.23	&	1.31	\\
SiO	&	2.99	&	2.95	&	3.11	\\
SO$_2$	&	1.54	&	1.52	&	1.63	\\
SO-triplet	&	1.40	&	1.40	&	1.56	\\
\end{tabular}
\end{ruledtabular}
\end{center}
\end{table*}

\begin{table*}[h]
\begin{center}
\caption{S22 weak interaction energies in kcal/mol.}
\label{stab:s22}
\begin{ruledtabular}
\begin{tabular}{crrr}
System & SCAN	&	rSCAN	& CCSD(T)\footnote{Reference [\onlinecite{B600027D}]} 	\\ 
\colrule
2-pyridoxine--2-aminopyridine	&	17.0	&	17.1	&	17	\\
Adenine--thymine stack	&	8.7	&	8.5	&	11.66	\\
Adenine--thymine WC	&	16.0	&	16.2	&	16.74	\\
Ammonia dimer	&	3.2	&	3.2	&	3.17	\\
Benzene--ammonia	&	2.1	&	2.0	&	2.32	\\
Benzene dimer C2h	&	1.0	&	0.9	&	2.62	\\
Benzene dimer C2v	&	1.5	&	1.5	&	2.71	\\
Benzene--HCN	&	4.2	&	4.1	&	4.55	\\
Benzene--methane	&	0.9	&	0.9	&	1.45	\\
Benzene--water	&	3.4	&	3.3	&	3.29	\\
Ethene dimer	&	1.2	&	1.1	&	1.5	\\
Ethene--ethyne	&	1.4	&	1.4	&	1.51	\\
Formamide dimer	&	16.6	&	16.7	&	16.12	\\
Formic acid dimer	&	21.0	&	20.9	&	18.8	\\
Indole--benzene stack	&	2.1	&	1.9	&	4.59	\\
Indole--benzene T-shape	&	4.2	&	4.1	&	5.62	\\
Methane dimer	&	0.4	&	0.4	&	0.53	\\
Phenol dimer	&	6.0	&	5.9	&	7.09	\\
Pyrazine dimer	&	2.7	&	2.5	&	4.2	\\
Uracil dimer HB	&	20.5	&	20.6	&	20.69	\\
Uracil dimer stack	&	8.1	&	7.9	&	9.74	\\
Water dimer	&	5.5	&	5.5	&	5.02	\\
\colrule
\end{tabular}
\end{ruledtabular}
\end{center}
\end{table*} 

\begin{table*}[h]
\begin{center}
\caption{BH76 barrier heights in kcal/mol. The forward and reverse barriers are shown.}
\label{stab:bh76}
\begin{ruledtabular}
\begin{tabular}{crrrr}
Reaction & Direction & SCAN	&	rSCAN	& Ref.\footnote{Reference [\onlinecite{doi:10.1021/jp045141s}]} 	\\ 
\colrule
H + HCl $\rightarrow$ H$_2$ + Cl   	    &	Forward	&	-1.4	&	-0.1	&	5.7	\\
	                                    &	Reverse	&	0.1	&	-0.3	&	8.7	\\
OH + H$_2$ $\rightarrow$ H$_2$O + H	    &	Forward	&	-2.1	&	-2.6	&	5.1	\\
                        	            &	Reverse	&	11.1	&	13.2	&	21.2	\\
CH$_3$ + H$_2$ $\rightarrow$ CH$_4$ + H	&	Forward	&	7.2	&	6.9	&	12.1	\\
	                                    &	Reverse	&	7.0	&	8.0	&	15.3	\\
OH + CH$_4$ $\rightarrow$ H$_2$O + CH$_3$	&	Forward	&	-1.6	&	-1.9	&	6.7	\\
	                                        &	Reverse	&	11.8	&	12.7	&	19.6	\\
H + H$_2$ $\rightarrow$ H$_2$ + H	        &	Forward	&	2.4	&	2.3	&	9.6	\\
	                                        &	Reverse	&	2.4	&	2.3	&	9.6	\\
OH + NH$_3$ $\rightarrow$ H$_2$O + NH$_2$	&	Forward	&	-7.4	&	-7.9	&	3.2	\\
	                                        &	Reverse	&	3.2	&	3.3	&	12.7	\\
HCl + CH$_3$ $\rightarrow$ CH$_4$ + Cl	    &	Forward	&	-3.1	&	-3.3	&	1.7	\\
	                                        &	Reverse	&	-1.7	&	-2.3	&	7.9	\\
OH + C$_2$H$_6$ $\rightarrow$ H$_2$O + C$_2$H$_5$	&	Forward	&	-4.8	&	-5.3	&	3.4	\\
	                                                &	Reverse	&	13.0	&	14.0	&	19.9	\\
F + H$_2$ $\rightarrow$ HF + H	            &	Forward	&	-7.7	&	-8.2	&	1.8	\\
	                                        &	Reverse	&	22.2	&	24.8	&	33.4	\\
O + CH$_4$ $\rightarrow$ OH + CH$_3$	    &	Forward	&	2.2	&	2.1	&	13.7	\\
	                                        &	Reverse	&	3.3	&	2.9	&	8.1	\\
H + PH$_3$ $\rightarrow$ H$_2$ + PH$_2$	&	Forward	&	-3.2	&	-3.4	&	3.1	\\
	                                    &	Reverse	&	19.5	&	19.3	&	23.2	\\
H + HO $\rightarrow$ H$_2$ + O	            &	Forward	&	3.2	&	3.0	&	10.7	\\
	                                        &	Reverse	&	2.1	&	1.1	&	13.1	\\
H + H$_2$S $\rightarrow$ H$_2$ + HS	    &	Forward	&	-2.7	&	-2.7	&	3.5	\\
	                                    &	Reverse	&	11.1	&	10.1	&	17.3	\\
O + HCl $\rightarrow$ OH + Cl	            &	Forward	&	-4.0	&	-5.0	&	9.8	\\
	                                        &	Reverse	&	-1.5	&	-3.2	&	10.4	\\
CH$_3$ + NH$_2$ $\rightarrow$ CH$_4$ + NH	&	Forward	&	4.5	&	3.9	&	8	\\
	                                        &	Reverse	&	12.5	&	12.8	&	22.4	\\
C$_2$H$_5$ + NH$_2$ $\rightarrow$ C$_2$H$_6$ + NH	&	Forward	&	6.0	&	5.5	&	7.5	\\
	                                                &	Reverse	&	9.5	&	9.6	&	18.3	\\
NH$_2$ + C$_2$H$_6$ $\rightarrow$ NH$_3$ + C$_2$H$_5$	&	Forward	&	4.8	&	4.5	&	10.4	\\
                                        	            &	Reverse	&	12.0	&	12.6	&	17.4	\\
NH$_2$ + CH$_4$ $\rightarrow$ NH$_3$ + CH$_3$	&	Forward	&	7.7	&	7.6	&	14.5	\\
	                                            &	Reverse	&	10.4	&	11.0	&	17.8	\\
s-trans cis-C$_5$H$_8$ $\rightarrow$ s-trans cis-C$_5$H$_8$	&	Forward	&	33.6	&	32.5	&	38.4	\\
	                                                        &	Reverse	&	33.6	&	32.5	&	38.4	\\
\end{tabular}
\end{ruledtabular}
\end{center}
\end{table*} 

\begin{table*}[h]
\begin{center}
\caption*{BH76 continued.}
\begin{ruledtabular}
\begin{tabular}{crrrr}
H + N$_2$O $\rightarrow$ OH + N$_2$	&	Forward	&	18.7	&	19.2	&	18.14	\\
	                                &	Reverse	&	66.1	&	62.6	&	83.22	\\
H + FH $\rightarrow$ HF + H	        &	Forward	&	38.3	&	38.2	&	42.18	\\
	                                &	Reverse	&	38.3	&	38.2	&	42.18	\\
H + ClH $\rightarrow$ HCl + H	        &	Forward	&	19.5	&	19.3	&	18	\\
	                                    &	Reverse	&	19.5	&	19.3	&	18	\\
H + FCH$_3$ $\rightarrow$ HF + CH$_3$	&	Forward	&	29.2	&	28.3	&	30.38	\\
	                                    &	Reverse	&	46.3	&	46.4	&	57.02	\\
H + F$_2$ $\rightarrow$ HF + F	        &	Forward	&	-1.6	&	-0.5	&	2.27	\\
	                                    &	Reverse	&	88.6	&	89.2	&	106.18	\\
CH$_3$ + FCl $\rightarrow$ CH$_3$F + Cl	&	Forward	&	-5.1	&	-4.7	&	7.43	\\
	                                    &	Reverse	&	45.9	&	45.4	&	60.17	\\
F$^-$ + CH$_3$F $\rightarrow$ FCH$_3$ + F$^-$	&	Forward	&	-8.3	&	-7.8	&	-0.34	\\
	                                            &	Reverse	&	-8.3	&	-7.8	&	-0.34	\\
F$^-$...CH$_3$F $\rightarrow$ FCH$_3$...F$^-$	    &	Forward	&	7.5	&	7.9	&	13.38	\\
	                                                &	Reverse	&	7.5	&	7.9	&	13.38	\\
Cl$^-$ + CH$_3$Cl $\rightarrow$ ClCH$_3$ + Cl$^-$	&	Forward	&	-6.1	&	-4.6	&	3.1	\\
	                                                &	Reverse	&	-6.1	&	-4.6	&	3.1	\\
Cl$^-$...CH$_3$Cl $\rightarrow$ ClCH$_3$...Cl$^-$	&	Forward	&	6.1	&	7.2	&	13.61	\\
	                                                &	Reverse	&	6.1	&	7.2	&	13.61	\\
F$^-$ + CH$_3$Cl $\rightarrow$ FCH$_3$ + Cl$^-$	&	Forward	&	-21.7	&	-20.5	&	-12.54	\\
                                	            &	Reverse	&	14.3	&	14.2	&	20.11	\\
F$^-$...CH$_3$Cl $\rightarrow$ FCH$_3$...Cl$^-$ &	Forward	&	-2.1	&	-1.2	&	2.89	\\
	                                            &	Reverse	&	24.4	&	24.2	&	29.62	\\
OH$^-$ + CH$_3$F $\rightarrow$ HOCH$_3$ + F$^-$	&	Forward	&	-11.2	&	-10.7	&	-2.78	\\
                                	            &	Reverse	&	9.8	&	10.8	&	17.33	\\
OH$^-$...CH$_3$F $\rightarrow$ HOCH$_3$...F$^-$	&	Forward	&	3.9	&	4.3	&	10.96	\\
	                                            &	Reverse	&	43.9	&	45.1	&	47.2	\\
H + N$_2$ $\rightarrow$ HN$_2$	        &	Forward	&	13.9	&	13.7	&	14.69	\\
                            	        &	Reverse	&	9.8	&	9.3	&	10.72	\\
H + CO $\rightarrow$ HCO               	&	Forward	&	5.9	&	6.1	&	3.17	\\
                            	        &	Reverse	&	24.2	&	24.2	&	22.68	\\
H + C$_2$H$_4$ $\rightarrow$ CH$_3$CH$_2$	    &	Forward	&	5.2	&	6.4	&	1.72	\\
	                                            &	Reverse	&	43.3	&	43.2	&	41.75	\\
CH$_3$ + C$_2$H$_4$ $\rightarrow$ CH$_3$CH$_2$CH$_2$	&	Forward	&	0.7	&	2.1	&	6.85	\\
	                                        &	Reverse	&	31.0	&	31.9	&	32.97	\\
HCN $\rightarrow$ HNC	        &	Forward	&	46.1	&	46.1	&	48.16	\\
	                            &	Reverse	&	32.0	&	31.6	&	33.11	\\
\end{tabular}
\end{ruledtabular}
\end{center}
\end{table*} 

\clearpage
\section{Infrared and Raman spectra of water cluster}

\begin{table*}[h]
\begin{center}
\caption{Infrared and Raman spectra of water monomer. Frequencies (cm$^{-1})$, IR intensities (km mol$^{-1}$), and Raman intensities (\AA$^4$ u$^{-1}$) and depolarization ratio are shown.}
\label{tab:spectra1}
\begin{ruledtabular}
\begin{tabular}{crrrcrrrr}
\multicolumn{4}{c}{SCAN}	&	\multicolumn{4}{c}{rSCAN}	& CCSD(T)\footnote{Reference \cite{doi:10.1063/1.4820448}} 	\\ \cmidrule(l){1-4} \cmidrule(l){5-8} \cmidrule(l){9-9}
Freq. &	IR 	&	Raman &	Depol.	&	Freq. &	IR &	Raman 	&	Depol.	&	Freq. 	\\  \cmidrule(l){1-1} \cmidrule(l){2-2} \cmidrule(l){3-3} \cmidrule(l){4-4} \cmidrule(l){5-5} \cmidrule(l){6-6} \cmidrule(l){7-7} \cmidrule(l){8-8} \cmidrule(l){9-9}
1640	&	70.77	&	0.66	&	0.72	&	1636	&	71.72	&	0.66	&	0.71	&	1638.1	\\
3802	&	2.96	&	100.86	&	0.05	&	3803	&	3.36	&	100.12	&	0.05	&	3786.8	\\
3909	&	54.50	&	25.46	&	0.75	&	3911	&	55.89	&	25.03	&	0.75	&	3904.5	\\
\end{tabular}
\end{ruledtabular}
\end{center}
\end{table*}

\begin{table*}[h]
\begin{center}
\caption{Infrared and Raman spectra of water dimer. Frequencies (cm$^{-1})$, IR intensities (km mol$^{-1}$), and Raman intensities (\AA$^4$ u$^{-1}$) and depolarization ratio are shown.}
\label{tab:spectra1}
\begin{ruledtabular}
\begin{tabular}{crrrcrrrr}
\multicolumn{4}{c}{SCAN}	&	\multicolumn{4}{c}{rSCAN}	& CCSD(T)\footnote{Reference \cite{doi:10.1063/1.4820448}} 	\\ \cmidrule(l){1-4} \cmidrule(l){5-8} \cmidrule(l){9-9}
Freq. &	IR 	&	Raman &	Depol.	&	Freq. &	IR &	Raman 	&	Depol.	&	Freq. 	\\  \cmidrule(l){1-1} \cmidrule(l){2-2} \cmidrule(l){3-3} \cmidrule(l){4-4} \cmidrule(l){5-5} \cmidrule(l){6-6} \cmidrule(l){7-7} \cmidrule(l){8-8} \cmidrule(l){9-9}
94	&	124.45	&	0.04	&	0.73	&	105	&	154.49	&	0.05	&	0.74	&	132.6	\\
147	&	47.70	&	0.05	&	0.74	&	163	&	22.31	&	0.03	&	0.71	&	145.3	\\
171	&	118.93	&	0.14	&	0.66	&	164	&	125.78	&	0.15	&	0.66	&	146.1	\\
203	&	173.20	&	0.06	&	0.54	&	201	&	161.80	&	0.04	&	0.58	&	183.6	\\
388	&	43.82	&	0.15	&	0.68	&	382	&	46.30	&	0.12	&	0.71	&	355.2	\\
649	&	89.10	&	0.34	&	0.74	&	653	&	88.74	&	0.32	&	0.74	&	629.7	\\
1639	&	85.86	&	0.92	&	0.75	&	1635	&	87.59	&	0.88	&	0.74	&	1639.6	\\
1664	&	36.35	&	0.72	&	0.30	&	1660	&	37.28	&	0.69	&	0.31	&	1658.8	\\
3654	&	386.50	&	151.68	&	0.14	&	3659	&	397.39	&	170.27	&	0.15	&	3712.1	\\
3790	&	8.85	&	83.27	&	0.05	&	3799	&	8.71	&	80.66	&	0.05	&	3782.5	\\
3876	&	74.56	&	51.35	&	0.26	&	3885	&	75.17	&	50.81	&	0.26	&	3875.5	\\
3894	&	78.12	&	23.97	&	0.75	&	3903	&	79.83	&	23.69	&	0.75	&	3896.1	\\
\end{tabular}
\end{ruledtabular}
\end{center}
\end{table*}  

\begin{table*}[h]
\begin{center}
\caption{Infrared and Raman spectra of water trimer. Frequencies (cm$^{-1})$, IR intensities (km mol$^{-1}$), and Raman intensities (\AA$^4$ u$^{-1}$) and depolarization ratio are shown.}
\label{tab:spectra1}
\begin{ruledtabular}
\begin{tabular}{crrrcrrrr}
\multicolumn{4}{c}{SCAN}	&	\multicolumn{4}{c}{rSCAN}	& CCSD(T)\footnote{Reference \cite{doi:10.1063/1.4820448}} 	\\ \cmidrule(l){1-4} \cmidrule(l){5-8} \cmidrule(l){9-9}
Freq. &	IR 	&	Raman &	Depol.	&	Freq. &	IR &	Raman 	&	Depol.	&	Freq. 	\\  \cmidrule(l){1-1} \cmidrule(l){2-2} \cmidrule(l){3-3} \cmidrule(l){4-4} \cmidrule(l){5-5} \cmidrule(l){6-6} \cmidrule(l){7-7} \cmidrule(l){8-8} \cmidrule(l){9-9}
190	&	102.95	&	0.04	&	0.71	&	200	&	37.87	&	0.03	&	0.71	&	157	\\
197	&	44.45	&	0.10	&	0.62	&	204	&	25.98	&	0.05	&	0.56	&	170	\\
213	&	57.00	&	0.19	&	0.74	&	220	&	71.17	&	0.30	&	0.70	&	183	\\
214	&	79.27	&	0.21	&	0.73	&	224	&	148.95	&	0.15	&	0.75	&	190.3	\\
240	&	7.45	&	0.27	&	0.18	&	239	&	5.47	&	0.26	&	0.18	&	216.7	\\
250	&	42.44	&	0.18	&	0.63	&	264	&	43.85	&	0.19	&	0.66	&	234.1	\\
375	&	62.91	&	0.93	&	0.34	&	385	&	76.09	&	0.72	&	0.41	&	334.5	\\
388	&	51.73	&	0.49	&	0.62	&	396	&	30.54	&	0.63	&	0.49	&	343.5	\\
481	&	122.23	&	0.49	&	0.28	&	488	&	124.93	&	0.53	&	0.25	&	434.1	\\
636	&	186.88	&	0.76	&	0.56	&	638	&	186.37	&	0.59	&	0.63	&	558.7	\\
719	&	290.05	&	0.30	&	0.46	&	724	&	291.11	&	0.27	&	0.57	&	650.1	\\
953	&	8.74	&	0.52	&	0.74	&	958	&	9.02	&	0.49	&	0.73	&	850.8	\\
1649	&	57.25	&	1.28	&	0.65	&	1647	&	52.22	&	1.29	&	0.63	&	1647.8	\\
1651	&	81.78	&	0.90	&	0.53	&	1649	&	89.09	&	0.88	&	0.52	&	1650.1	\\
1674	&	14.87	&	0.98	&	0.75	&	1672	&	14.12	&	0.98	&	0.75	&	1671.7	\\
3429	&	29.13	&	304.92	&	0.06	&	3427	&	26.38	&	300.24	&	0.06	&	3596.5	\\
3515	&	754.03	&	37.68	&	0.74	&	3514	&	759.62	&	37.20	&	0.74	&	3647.8	\\
3537	&	683.55	&	47.93	&	0.54	&	3533	&	687.80	&	38.67	&	0.59	&	3655	\\
3864	&	78.86	&	53.18	&	0.22	&	3865	&	87.49	&	46.96	&	0.25	&	3865.7	\\
3865	&	77.77	&	51.03	&	0.43	&	3866	&	70.63	&	62.53	&	0.32	&	3869.9	\\
3867	&	42.63	&	114.89	&	0.08	&	3869	&	45.91	&	109.97	&	0.09	&	3871.4	\\
\end{tabular}
\end{ruledtabular}
\end{center}
\end{table*}

\begin{table*}[h]
\begin{center}
\caption{Infrared and Raman spectra of water tetramer. Frequencies (cm$^{-1})$, IR intensities (km mol$^{-1}$), and Raman intensities (\AA$^4$ u$^{-1}$) and depolarization ratio are shown.}
\label{tab:spectra1}
\begin{ruledtabular}
\begin{tabular}{crrrcrrrr}
\multicolumn{4}{c}{SCAN}	&	\multicolumn{4}{c}{rSCAN}	& CCSD(T)\footnote{Reference \cite{doi:10.1063/1.4820448}} 	\\ \cmidrule(l){1-4} \cmidrule(l){5-8} \cmidrule(l){9-9}
Freq. &	IR 	&	Raman &	Depol.	&	Freq. &	IR &	Raman 	&	Depol.	&	Freq. 	\\  \cmidrule(l){1-1} \cmidrule(l){2-2} \cmidrule(l){3-3} \cmidrule(l){4-4} \cmidrule(l){5-5} \cmidrule(l){6-6} \cmidrule(l){7-7} \cmidrule(l){8-8} \cmidrule(l){9-9}
63	&	0.15	&	0.17	&	0.67	&	54	&	0.01	&	0.12	&	0.73	&	48.8	\\
77	&	2.91	&	0.30	&	0.74	&	85	&	1.90	&	0.39	&	0.75	&	76.1	\\
222	&	0.16	&	0.24	&	0.11	&	223	&	0.10	&	0.22	&	0.12	&	192.1	\\
236	&	38.03	&	0.96	&	0.75	&	234	&	37.55	&	0.81	&	0.75	&	208.2	\\
251	&	130.83	&	0.04	&	0.75	&	263	&	37.78	&	0.05	&	0.75	&	230.8	\\
260	&	47.33	&	0.06	&	0.74	&	265	&	20.32	&	0.05	&	0.75	&	230.8	\\
277	&	146.05	&	0.03	&	0.66	&	283	&	234.89	&	0.00	&	0.75	&	248.7	\\
282	&	216.38	&	0.00	&	0.34	&	286	&	248.59	&	0.00	&	0.73	&	248.7	\\
287	&	9.42	&	0.03	&	0.60	&	289	&	3.94	&	0.05	&	0.75	&	254.3	\\
316	&	1.50	&	0.22	&	0.36	&	327	&	0.31	&	0.20	&	0.34	&	283.5	\\
433	&	0.20	&	0.19	&	0.29	&	442	&	0.02	&	0.21	&	0.29	&	391	\\
461	&	19.63	&	0.78	&	0.74	&	478	&	15.73	&	0.80	&	0.75	&	421	\\
481	&	38.75	&	0.56	&	0.75	&	493	&	37.35	&	0.52	&	0.75	&	437.9	\\
484	&	37.65	&	0.62	&	0.75	&	494	&	38.02	&	0.59	&	0.75	&	437.9	\\
807	&	132.49	&	0.62	&	0.73	&	811	&	136.70	&	0.69	&	0.75	&	730.5	\\
891	&	159.02	&	0.41	&	0.75	&	898	&	163.54	&	0.40	&	0.75	&	800.2	\\
900	&	161.73	&	0.37	&	0.74	&	900	&	161.87	&	0.41	&	0.75	&	800.2	\\
1074	&	0.02	&	0.68	&	0.01	&	1078	&	0.00	&	0.42	&	0.08	&	971	\\
1649	&	83.66	&	0.66	&	0.75	&	1647	&	84.91	&	0.64	&	0.75	&	1653	\\
1664	&	42.80	&	0.54	&	0.75	&	1663	&	42.73	&	0.53	&	0.75	&	1666.6	\\
1665	&	42.21	&	0.54	&	0.74	&	1664	&	42.67	&	0.55	&	0.75	&	1666.6	\\
1697	&	0.05	&	1.75	&	0.20	&	1696	&	0.00	&	1.91	&	0.19	&	1693.3	\\
3161	&	0.21	&	360.22	&	0.08	&	3160	&	0.23	&	357.14	&	0.08	&	3446.4	\\
3299	&	1886.40	&	2.13	&	0.73	&	3299	&	1893.09	&	2.10	&	0.74	&	3526.6	\\
3303	&	1917.26	&	2.27	&	0.72	&	3300	&	1891.83	&	2.38	&	0.73	&	3526.6	\\
3354	&	26.31	&	105.31	&	0.75	&	3353	&	26.84	&	106.52	&	0.75	&	3559.1	\\
3860	&	67.22	&	48.06	&	0.35	&	3861	&	69.75	&	44.94	&	0.34	&	3860.9	\\
3861	&	45.97	&	95.66	&	0.08	&	3862	&	73.59	&	39.86	&	0.34	&	3861.4	\\
3862	&	71.11	&	43.90	&	0.45	&	3863	&	45.34	&	97.98	&	0.11	&	3861.4	\\
3864	&	45.77	&	95.13	&	0.07	&	3865	&	46.19	&	95.84	&	0.09	&	3861.5	\\
\end{tabular}
\end{ruledtabular}
\end{center}
\end{table*}  

\begingroup
\begin{table*}[h]
\begin{center}
\caption{Infrared and Raman spectra of water pentamer. Frequencies (cm$^{-1})$, IR intensities (km mol$^{-1}$), and Raman intensities (\AA$^4$ u$^{-1}$) and depolarization ratio are shown.}
\label{tab:spectra1}
\begin{ruledtabular}
\begin{tabular}{crrrcrrrr}
\multicolumn{4}{c}{SCAN}	&	\multicolumn{4}{c}{rSCAN}	& CCSD(T)\footnote{Reference \cite{doi:10.1063/1.4820448}} 	\\ \cmidrule(l){1-4} \cmidrule(l){5-8} \cmidrule(l){9-9}
Freq. &	IR 	&	Raman &	Depol.	&	Freq. &	IR &	Raman 	&	Depol.	&	Freq. 	\\  \cmidrule(l){1-1} \cmidrule(l){2-2} \cmidrule(l){3-3} \cmidrule(l){4-4} \cmidrule(l){5-5} \cmidrule(l){6-6} \cmidrule(l){7-7} \cmidrule(l){8-8} \cmidrule(l){9-9}
30	&	4.38	&	0.02	&	0.75	&	30	&	4.57	&	0.02	&	0.74	&	22.5	\\
44	&	0.06	&	0.18	&	0.62	&	43	&	0.03	&	0.20	&	0.61	&	41.4	\\
68	&	0.10	&	0.22	&	0.74	&	67	&	0.15	&	0.25	&	0.75	&	60.6	\\
78	&	1.84	&	0.29	&	0.66	&	71	&	1.45	&	0.26	&	0.75	&	63.5	\\
189	&	2.21	&	0.21	&	0.22	&	191	&	0.91	&	0.22	&	0.15	&	179.2	\\
203	&	31.05	&	0.59	&	0.74	&	206	&	31.14	&	0.60	&	0.75	&	190.1	\\
213	&	31.08	&	0.62	&	0.74	&	219	&	39.08	&	0.59	&	0.75	&	196	\\
238	&	115.26	&	0.17	&	0.70	&	244	&	95.34	&	0.17	&	0.73	&	226.6	\\
255	&	4.33	&	0.01	&	0.73	&	258	&	6.82	&	0.01	&	0.71	&	232.3	\\
267	&	146.94	&	0.03	&	0.74	&	266	&	148.96	&	0.04	&	0.71	&	236.9	\\
286	&	270.53	&	0.09	&	0.29	&	288	&	276.54	&	0.05	&	0.59	&	265.1	\\
321	&	82.37	&	0.18	&	0.74	&	324	&	76.91	&	0.17	&	0.75	&	291.8	\\
330	&	11.90	&	0.06	&	0.75	&	331	&	19.75	&	0.04	&	0.73	&	295	\\
336	&	1.11	&	0.04	&	0.70	&	338	&	1.52	&	0.02	&	0.72	&	297.7	\\
449	&	48.52	&	0.84	&	0.39	&	453	&	50.97	&	0.62	&	0.46	&	400.5	\\
463	&	14.93	&	0.31	&	0.22	&	467	&	10.20	&	0.30	&	0.16	&	420.7	\\
490	&	15.64	&	0.81	&	0.74	&	495	&	15.78	&	0.75	&	0.74	&	443.9	\\
500	&	17.78	&	0.79	&	0.73	&	507	&	19.11	&	0.79	&	0.73	&	456.7	\\
559	&	62.28	&	0.99	&	0.35	&	561	&	62.38	&	0.89	&	0.40	&	503.1	\\
771	&	23.74	&	1.31	&	0.65	&	778	&	24.16	&	1.32	&	0.70	&	702.5	\\
848	&	128.11	&	1.18	&	0.69	&	852	&	126.67	&	1.07	&	0.74	&	766.2	\\
934	&	130.17	&	0.62	&	0.69	&	941	&	130.08	&	0.59	&	0.75	&	843.9	\\
955	&	124.66	&	0.26	&	0.73	&	960	&	123.33	&	0.22	&	0.73	&	860	\\
1063	&	8.15	&	0.07	&	0.19	&	1068	&	8.54	&	0.15	&	0.11	&	963.7	\\
1649	&	83.44	&	0.17	&	0.69	&	1647	&	83.94	&	0.18	&	0.74	&	1657.6	\\
1662	&	20.27	&	0.72	&	0.58	&	1660	&	19.34	&	0.72	&	0.59	&	1667.9	\\
1672	&	58.45	&	0.55	&	0.37	&	1671	&	61.06	&	0.49	&	0.42	&	1675.2	\\
1696	&	34.66	&	0.48	&	0.05	&	1695	&	36.19	&	0.44	&	0.04	&	1694.4	\\
1705	&	4.22	&	0.62	&	0.63	&	1703	&	3.96	&	0.73	&	0.49	&	1701.4	\\
3095	&	50.75	&	495.91	&	0.08	&	3097	&	48.68	&	493.31	&	0.08	&	3413	\\
3223	&	2923.51	&	4.41	&	0.74	&	3224	&	2857.86	&	4.04	&	0.74	&	3482.9	\\
3235	&	2624.40	&	13.29	&	0.38	&	3237	&	2630.15	&	13.05	&	0.38	&	3490.2	\\
3297	&	93.02	&	74.00	&	0.75	&	3298	&	79.76	&	73.89	&	0.74	&	3529.6	\\
3317	&	113.31	&	79.85	&	0.72	&	3317	&	120.20	&	75.18	&	0.73	&	3535.5	\\
3861	&	50.36	&	53.55	&	0.28	&	3867	&	56.87	&	60.48	&	0.19	&	3859.1	\\
3862	&	67.18	&	61.22	&	0.15	&	3870	&	66.32	&	49.40	&	0.24	&	3861.1	\\
3862	&	50.73	&	57.29	&	0.18	&	3871	&	49.89	&	88.56	&	0.12	&	3862.9	\\
3864	&	52.53	&	87.50	&	0.12	&	3873	&	40.97	&	23.24	&	0.71	&	3862.9	\\
3867	&	57.65	&	108.78	&	0.08	&	3875	&	67.76	&	142.57	&	0.05	&	3865.7	\\
\end{tabular}
\end{ruledtabular}
\end{center}
\end{table*}  
\endgroup

\clearpage
\section{FLOSIC-rSCAN calculations}

\begin{table*}[h]
\begin{center}
\caption{Atoms: total energies (in Ha).}
\label{stab:flosicatoms}
\begin{ruledtabular}
\begin{tabular}{crrr}
Z & rSCAN	&	SIC-rSCAN	& E$_\text{Accu}$ (Ref. [\onlinecite{PhysRevA.47.3649}]) 	\\ 
\colrule
1	&	-0.500	&	-0.500	&	-0.5	\\
2	&	-2.905	&	-2.900	&	-2.90	\\
3	&	-7.480	&	-7.474	&	-7.48	\\
4	&	-14.650	&	-14.643	&	-14.67	\\
5	&	-24.641	&	-24.628	&	-24.65	\\
6	&	-37.841	&	-37.813	&	-37.85	\\
7	&	-54.594	&	-54.538	&	-54.59	\\
8	&	-75.076	&	-74.997	&	-75.07	\\
9	&	-99.752	&	-99.640	&	-99.73	\\
10	&	-128.963	&	-128.799	&	-128.94	\\
11	&	-162.286	&	-162.100	&	-162.25	\\
12	&	-200.082	&	-199.874	&	-200.05	\\
13	&	-242.383	&	-242.147	&	-242.35	\\
14	&	-289.404	&	-289.131	&	-289.36	\\
15	&	-341.311	&	-340.993	&	-341.26	\\
16	&	-398.164	&	-397.807	&	-398.11	\\
17	&	-460.208	&	-459.802	&	-460.15	\\
18	&	-527.606	&	-527.141	&	-527.54	\\
\colrule
19	&	-599.981	&	-599.467	&		\\
20	&	-677.627	&	-677.061	&		\\
21	&	-760.692	&	-760.077	&		\\
22	&	-849.446	&	-848.791	&		\\
23	&	-944.011	&	-943.254	&		\\
24	&	-1044.596	&	-1043.686	&		\\
25	&	-1151.143	&	-1150.196	&		\\
26	&	-1263.849	&	-1262.832	&		\\
27	&	-1382.936	&	-1381.869	&		\\
28	&	-1508.506	&	-1507.310	&		\\
29	&	-1640.748	&	-1639.332	&		\\
30	&	-1779.669	&	-1778.237	&		\\
31	&	-1925.102	&	-1923.629	&		\\
32	&	-2077.232	&	-2075.698	&		\\
33	&	-2236.145	&	-2234.541	&		\\
34	&	-2401.837	&	-2400.160	&		\\
35	&	-2574.471	&	-2572.710	&		\\
36	&	-2754.143	&	-2752.290	&		\\
\end{tabular}
\end{ruledtabular}
\end{center}
\end{table*}

\begin{table*}[h]
\begin{center}
\caption{Atoms: ionization potentials (in eV).}
\label{stab:ips}
\begin{ruledtabular}
\begin{tabular}{crrr}
Z & rSCAN	&	SIC-rSCAN	& Expt.\footnote{Reference \cite{NIST_ASD}} 	\\ 
\colrule
2	&	24.624	&	24.483	&	24.587	\\
3	&	5.400	&	5.374	&	5.392	\\
4	&	8.802	&	8.820	&	9.323	\\
5	&	8.788	&	8.732	&	8.298	\\
6	&	11.716	&	11.437	&	11.26	\\
7	&	14.889	&	14.308	&	14.534	\\
8	&	13.707	&	13.431	&	13.618	\\
9	&	17.609	&	17.005	&	17.423	\\
10	&	21.633	&	20.589	&	21.565	\\
11	&	5.180	&	5.140	&	5.139	\\
12	&	7.393	&	7.395	&	7.646	\\
13	&	6.181	&	6.180	&	5.986	\\
14	&	8.341	&	8.215	&	8.152	\\
15	&	10.670	&	10.441	&	10.487	\\
16	&	10.345	&	10.419	&	10.36	\\
17	&	13.045	&	12.962	&	12.968	\\
18	&	15.879	&	15.646	&	15.76	\\
\colrule
19	&	4.282	&	4.472	&	4.341	\\
20	&	5.839	&	6.089	&	6.113	\\
21	&	6.241	&	7.054	&	6.561	\\
22	&	6.984	&	8.324	&	6.828	\\
23	&	7.126	&	7.070	&	6.746	\\
24	&	7.312	&	7.015	&	6.767	\\
25	&	6.908	&	7.133	&	7.434	\\
26	&	7.795	&	8.068	&	7.902	\\
27	&	8.396	&	8.279	&	7.881	\\
28	&	8.789	&	8.443	&	7.64	\\
29	&	8.090	&	7.488	&	7.726	\\
30	&	9.238	&	9.519	&	9.394	\\
31	&	6.156	&	6.560	&	5.999	\\
32	&	8.093	&	8.324	&	7.899	\\
33	&	10.134	&	10.423	&	9.789	\\
34	&	9.685	&	10.280	&	9.752	\\
35	&	11.913	&	12.379	&	11.814	\\
36	&	14.250	&	14.743	&	14	\\
\end{tabular}
\end{ruledtabular}
\end{center}
\end{table*}

\begin{table*}[h]
\begin{center}
\caption{Atoms: electron affinities (in eV).}
\label{stab:eas}
\begin{ruledtabular}
\begin{tabular}{crrr}
Z & rSCAN	&	SIC-rSCAN	& Expt.\footnote{Reference \cite{NIST_CCCBD}} 	\\ 
\colrule
1	&	0.725	&	0.510	&	0.754	\\
3	&	0.446	&	0.465	&	0.618	\\
5	&	0.608	&	0.167	&	0.280	\\
6	&	1.547	&	0.875	&	1.262	\\
8	&	1.573	&	0.637	&	1.462	\\
9	&	3.428	&	2.123	&	3.401	\\
11	&	0.457	&	0.480	&	0.548	\\
13	&	0.626	&	0.383	&	0.434	\\
14	&	1.570	&	1.277	&	1.390	\\
15	&	0.765	&	0.602	&	0.747	\\
16	&	2.157	&	1.843	&	2.077	\\
17	&	3.714	&	3.277	&	3.613	\\
\colrule
19	&	0.409	&	0.424	&	0.501	\\
22	&	1.008	&	-1.211	&	0.087	\\
29	&	1.178	&	1.061	&	1.236	\\
31	&	0.544	&	0.187	&	0.43	\\
32	&	1.540	&	1.274	&	1.233	\\
33	&	0.830	&	0.659	&	0.814	\\
34	&	2.135	&	1.806	&	2.021	\\
35	&	3.584	&	3.232	&	3.364	\\
\end{tabular}
\end{ruledtabular}
\end{center}
\end{table*}

\begin{table*}[h]
\begin{center}
\caption{AE6 atomization energies (in kcal/mol).}
\label{stab:atomization}
\begin{ruledtabular}
\begin{tabular}{crrr}
System & rSCAN	&	SIC-rSCAN	& Ref.\footnote{Reference \cite{doi:10.1021/jp035287b}} 	\\ 
\colrule
C$_2$O$_2$H$_2$	& 643.0	& 589.5	& 634.0 \\
CH$_3$CCH	& 710.2	& 678.3	& 705.1 \\
C$_4$H$_8$	& 1160.8 & 1134.0 & 1149.4 \\
S$_2$	    & 109.6	& 98.7	& 104.3 \\
SiH$_4$	& 322.4	& 326.9	& 325.0 \\
SiO	    & 188.8	& 157.8	& 193.1 \\
\end{tabular}
\end{ruledtabular}
\end{center}
\end{table*}

\begin{table*}[h]
\begin{center}
\caption{BH6 barrier heights (in kcal/mol).}
\label{stab:barrierheight}
\begin{ruledtabular}
\begin{tabular}{ccrrr}
Reaction & Direction & rSCAN	&	SIC-rSCAN	& Ref.\footnote{Reference \cite{doi:10.1021/jp035287b}} 	\\ 
\colrule
OH + CH$_4\rightarrow$ CH$_3$+ H$_2$O   &	Forward   &	-14.6   &	11.6	&   6.7 \\
&	Reverse   &	12.6	&   14.2	&   19.6 \\
H +OH$\rightarrow$ H$_2$ + O        &	Forward   &	2.1	    &   10.6	&   10.7 \\
&   Reverse   &	12.8	&   14.0	&   13.1 \\
H + H$_2$S$\rightarrow$ H$_2$+ HS	    &   Forward   &	-2.7	&   1.6	    &   3.6 \\
&   Reverse   &	4.3	    &   14.3	&   17.3 \\
\end{tabular}
\end{ruledtabular}
\end{center}
\end{table*}

\begin{table*}[h]
\begin{center}
\caption{SIE4$\times$4 dissociation energies and SIE11 reaction energies (in kcal/mol).}
\label{stab:dissociationandreaction}
\begin{ruledtabular}
\begin{tabular}{crrr}
Reaction & rSCAN	&	SIC-rSCAN	& Ref.\footnote{Reference \cite{doi:10.1021/ct900489g}} 	\\ 
\colrule
H$_2^+$	$\rightarrow$	H	$+$	H$^+$	&		&		&		\\ 
R/R$_e$ = 1.0					&	67.8	&	64.4	&	64.4	\\
R/R$_e$ = 1.25					&	64.9	&	58.9	&	58.9	\\
R/R$_e$ = 1.5					&	57.8	&	48.7	&	48.7	\\
R/R$_e$ = 1.75					&	50.8	&	38.2	&	38.3	\\
\colrule
He$_2^+$	$\rightarrow$	He	$+$	He$^+$	&		&		&		\\ 
R/R$_e$ = 1.0					&	74.4	&	56.5	&	56.9	\\
R/R$_e$ = 1.25					&	71.5	&	44.6	&	46.9	\\
R/R$_e$ = 1.5					&	63.4	&	27.5	&	31.3	\\
R/R$_e$ = 1.75					&	58.5	&	14.3	&	19.1	\\
\colrule
(NH$_3$)$_2^+$	$\rightarrow$	NH$_3$	$+$	NH$_3^+$	&		&		&		\\ 
R/R$_e$ = 1.0					&	43.4	&	36.3	&	35.9	\\
R/R$_e$ = 1.25					&	38.3	&	25.4	&	25.9	\\
R/R$_e$ = 1.5					&	30.9	&	11.6	&	13.4	\\
R/R$_e$ = 1.75					&	27.2	&	4.1	&	4.9	\\
\colrule
(H$_2$O)$_2^+$	$\rightarrow$	H$_2$O	$+$	H$_2$O$^+$	&		&		&		\\ 
R/R$_e$ = 1.0					&	52.9	&	36.3	&	39.7	\\
R/R$_e$ = 1.25					&	48.8	&	22.8	&	29.1	\\
R/R$_e$ = 1.5					&	42.7	&	11.9	&	16.9	\\
R/R$_e$ = 1.75					&	40.1	&	6.2	&	9.3	\\
\colrule
C$_4$H$_{10}^+$	$\rightarrow$	C$_2$H$_5$	$+$	C$_2$H$_5^+$	&	42.0	&	34.3	&	35.28	\\
(CH$_3$)$_2$CO$^+$	$\rightarrow$	CH$_3$	$+$	CH$_3$CO$^+$	&	30.1	&	40.5	&	22.57	\\
ClFCl	$\rightarrow$	ClClF			&	-22.3	&	-2.9	&	-1.01	\\
C$_2$H$_4$...F$_2$	$\rightarrow$	C$_2$H$_4$	$+$	F$_2$	&	2.5	&	0.5	&	1.08	\\
C$_6$H$_6$...Li	$\rightarrow$	Li	$+$	C$_6$H$_6$	&	7.7	&	12.1	&	9.5	\\
NH$_3$...ClF	$\rightarrow$	NH$_3$	$+$	ClF	&	17.1	&	12.0	&	10.5	\\
NaOMg	$\rightarrow$	MgO	$+$	Na	&	75.7	&	95.3	&	69.56	\\
FLiF	$\rightarrow$	Li	$+$	F$_2$	&	120.4	&	92.4	&	94.36	\\
\end{tabular}
\end{ruledtabular}
\end{center}
\end{table*}

\clearpage
\bibliography{bibtex_references}